\documentclass[preprint,12pt]{elsarticle}
\usepackage[top=1in, bottom=1in, left=1in, right=1in]{geometry}

\usepackage{amsmath,amssymb,amsfonts, bm}

\usepackage{array}
\usepackage{tabularx}
\usepackage{booktabs} 
\usepackage{multirow}
\usepackage{enumitem}
\usepackage{balance}
\usepackage{url}

\usepackage{graphicx}
\usepackage{stfloats}
\usepackage{subcaption}

\usepackage{textcomp}
\usepackage{verbatim}
\usepackage{lipsum}
\usepackage{xcolor}  
\usepackage{fontawesome5}
\usepackage{orcidlink}
\usepackage{calc}
\usepackage{algorithm}
\usepackage{algcompatible}

\newcommand{\mbf}[1]{\mathbf{#1}}
\newcommand{\blue}[1]{\textcolor{black}{#1}}  
\definecolor{belowb1}{RGB}{245, 245, 220}  

\usepackage[utf8]{inputenc}

\journal{Reliability Engineering and System Safety}

\begin{document}

\begin{frontmatter}

\title{Repulsive normalizing flow mixtures for adaptive importance sampling: reliability analysis of complex systems} 

\author[inst1]{Sara Helal\corref{cor1}}
\author[inst1]{Victor Elvira}

\cortext[cor1]{Corresponding author\\\hspace*{1.65em}E-mail address: S.Helal@sms.ed.ac.uk}

\affiliation[inst1]{
  organization={School of Mathematics, University of Edinburgh},
  city={Edinburgh},
  postcode={EH9 3FD},
  country={UK}
}
\date{} 
\begin{abstract}
Accurate rare-event estimation can be computationally expensive. Classical adaptive importance sampling (IS) schemes often rely on restrictive proposal families and can struggle under multiple failure modes. We propose FAMIS, a flow-based multiple importance sampling (MIS) framework that learns a nonuniform mixture of normalizing flow proposals for rare-event estimation. The method does not require presampled failure data or prior knowledge of the number, location, or geometry of the failure modes. Instead, it adaptively learns the {mixture} through sequential evaluations of the limit state function. To guide training toward the failure domain, FAMIS uses a smooth rare-event surrogate and a tempered target sequence. A defensive exploration mixture improves early-stage coverage, a Rao--Blackwellized update adapts the mixture weights, and a Jensen--Shannon repulsion term promotes separation and diversity among the base components. The final failure probability is computed with a deterministic-mixture MIS estimator. Numerical experiments demonstrate that FAMIS accurately approximates quasi-optimal IS densities with fewer training samples and model evaluations, providing stable variance reduction across complex reliability problems.
\end{abstract}

\begin{keyword}
Adaptive importance sampling; high-dimensional reliability analysis; Monte Carlo methods; multiple failure modes; normalizing flows;  non-Gaussian; rare events.
\end{keyword}

\end{frontmatter}

\section{Introduction}
\label{sec1}

Rare-event probability estimation is a fundamental problem in reliability analysis and uncertainty quantification. It has applications across many engineering domains, including structural and mechanical systems \cite{wei2019structural}, aerospace engineering \cite{morio2015estimation}, energy infrastructures \cite{cadini2017estimation}, and autonomous systems \cite{10711532BAI}.
In reliability problems, the system response is commonly described by a limit state function (LSF) \(S(\mathbf{x})\), where \(\mathbf{x}\) denotes the uncertain input variables. The rare-event probability $P_f$ is therefore the probability that random inputs fall inside the failure region $S(\mathbf{x})\leq 0$. When the failure region is disconnected or contains several separated failure domains, these areas are referred to as failure modes \cite{HELAL2026}.
The events of interest correspond to failures induced by extreme loads, rare environmental conditions, or unlikely but high-impact system configurations \cite{GUO2025111078}. These probabilities are often very small, typically in the range of $10^{-4}$ to $10^{-9}$ or lower, which makes their accurate estimation both mathematically and computationally challenging. In such regimes, crude Monte Carlo methods become prohibitively expensive, since an impractically large number of simulations is required to observe a sufficient number of failures.

To address this computational bottleneck, existing reliability estimation methods can be broadly grouped into three main families: approximation-based methods, surrogate-model methods, and sampling-based methods. Approximation-based techniques, such as first- and second-order reliability methods (FORM and SORM), replace the original LSF by a local approximation around a design point, also known as the most probable point \cite{keshtegar2017hybrid,gong2017first,lim2016post,zhao1999general}. These methods are attractive because of their low computational cost and their clear geometric interpretation. However, their accuracy can deteriorate when the LSF is strongly nonlinear, when the input dimension is large, or when the failure domain contains several separated design points. In such cases, a local approximation may provide an incomplete description of the geometry that drives the failure probability.

A second line of work relies on surrogate models, where the expensive LSF is replaced by a cheaper metamodel trained from selected evaluations. Popular choices include response surfaces \cite{lucia2763}, polynomial chaos expansions \cite{PCK0000870,schobi2015polynomial}, support vector machines \cite{HURTADO2004271}, adaptive Kriging models \cite{zhang2022moving,bichon2008efficient}, and neural networks \cite{BAO2021107778}. These approaches can substantially reduce the cost of reliability analysis when each evaluation of the physical model is expensive. Their performance, however, depends critically on the quality and placement of the training points, especially near the boundary of the failure region \cite{XIAO2020106852}. As a result, surrogate methods are often combined with sampling strategies to refine the approximation near rare-event regions and to reduce the bias that may arise when estimating very small probabilities.

Sampling-based methods form a third and more general family. A large body of variance-reduction methods, including directional sampling \cite{guo2020active,grooteman2011adaptive}, line sampling \cite{PRADLWARTER2007208}, subset simulation \cite{AU2001263}, importance sampling (IS) \cite{elvira2021advances}, adaptive importance sampling (AIS) \cite{cappe2008adaptive,martino2014adaptive}, subset adaptive importance sampling (SAIS) \cite{HELAL2026}, and population Monte Carlo (PMC) methods \cite{cappe2004population,elvira2022optimized,ELVIRA201777}. More recent developments construct sequences of intermediate distributions or failure thresholds to gradually move the sampling effort toward rare-event regions, including sequential IS schemes \cite{PAPAIOANNOU201666}, cross-entropy approaches \cite{geyer2019cross,uribe2021cross}, relaxation-based methods \cite{XIAN2024102393}, large-deviation-driven strategies \cite{22M1524758}, directional adaptive methods \cite{CHENG2023102291}, and fault-tree-informed adaptive IS for piecewise deterministic Markov processes \cite{22M1522838}.

Among these approaches, IS is particularly appealing because it preserves unbiasedness, or asymptotic consistency under standard self-normalized variants \cite{elvira2021advances}.
The basic principle is to generate samples from a so-called importance distribution (i.e., proposal) that better explores the failure domain and, consequently, reduces the variance of the estimator. Since these samples are no longer drawn from the original input distribution, each sample is assigned an importance weight that corrects for the change of sampling distribution. 
 Its efficiency is nevertheless governed by the proposal distribution. A proposal that misses part of the failure region may lead to unstable weights and large variance, whereas an overly spread proposal wastes samples in irrelevant regions. This difficulty becomes especially pronounced for nonlinear and complex failure domains. Therefore, a central question in modern adaptive IS is not merely how to reweight samples, but how to adapt proposal distributions that can learn the geometry of rare events without requiring a priori knowledge of the failure set \cite{elvira2019generalized}.

 {In recent years, the rare-event simulation literature has increasingly adopted machine learning-based deep generative models, particularly normalizing flows (NFs), to construct adaptive IS proposals. By transforming a simple base density into a flexible failure-focused proposal, NFs enable exact sampling and tractable density evaluation through change of variables. Several recent methods have explored this direction: flow-based generative samplers were trained to approximate the conditional rare-event distribution and then combined with IS in \cite{gibson2023flow}; REIN used NFs to learn quasi-optimal IS proposals for reliability estimation \cite{DASGUPTA2024109729}; and NOFIS constructed a sequence of NF-based proposals associated with nested subset events \cite{gao2024nofis}.  Related developments beyond classical reliability include FlowRES for rare transition path sampling~\cite{asghar2024flowres} and self-regularized NFs for rare-event modeling with limited failure data~\cite{dawson2025rare}.}

 Despite the flexibility of these flow-based samplers, most existing methods typically rely on a single global flow or multiple independent flows that do not coordinate their exploration of the state space. Consequently, maintaining component diversity, covering multiple disconnected failure modes, and stabilizing the adaptation of mixture proposals remain challenging, especially when the rare-event set is geometrically complex. Moreover, the quasi-optimal IS density over the failure set is typically highly non-Gaussian,
often exhibiting strong curvature and multiple separated modes induced by nonlinear limit state functions. 
Standard adaptive proposals based on mixtures of Gaussians may therefore require many local updates and components to approximate such geometry. Similarly, many sampling-based and surrogate methods rely on repeatedly obtaining elite or near-failure samples to
guide their adaptation.
This is an assumption that becomes weak when failures are extremely rare and no presampled failure data are available.
 \vspace{5pt}

\noindent\textbf{Contribution and novelty.} In this paper, we propose \textit{FAMIS}, an adaptive flow-based multiple importance sampling framework for rare-event probability estimation in multimodal, high-dimensional, and non-Gaussian reliability problems. The main contributions are summarized as follows:

\begin{itemize}

    \item We introduce a flow-based MIS framework in which multiple adaptive base proposals are transformed through a shared normalizing flow. The shared transport learns a global transformation that simplifies the rare-event target geometry, while the individual mixture components capture distinct local failure regions. 

    \item FAMIS requires neither presampled failures nor prior knowledge of the location or geometry of the failure modes. It learns proposals on-the-fly from LSF evaluations using a smooth surrogate and tempered targets, without requiring LSF derivatives, and is therefore applicable to black-box performance models.

    \item We use a nonuniform flow mixture with defensive exploration, combining flexible failure-focused proposals with a broad auxiliary distribution to improve coverage of poorly explored failure regions. Rao--Blackwellized weight adaptation then allocates sampling effort across components according to their contribution to the current target.

    \item We introduce a novel training objective based on a soft-weighted Kullback--Leibler (KL) loss to improve discovery of poorly covered or disconnected failure regions. The forward-KL component promotes mode discovery, while the reverse-KL component fits the proposal around captured failure modes.

\item We develop a Jensen--Shannon repulsion mechanism between the trainable mixture components that mitigates component collapse, thereby promoting proposal diversity and improving coverage of multimodal failure regions.

    \item We validate the performance of the method on challenging multimodal, nonlinear, non-Gaussian, and high-dimensional reliability problems.

\end{itemize}

 The remainder of this paper is organized as follows. Section~\ref{sec2} formulates the rare-event problem in the context of probabilistic reliability analysis. Section~\ref{sec3} reviews the main background concepts, including adaptive importance sampling and normalizing flows. Sections~\ref{sec4} and~\ref{sec5} present the proposed method, FAMIS, and its proposal adaptation, respectively. Section~\ref{sec6} discusses design choices and parameter tuning. Section~\ref{sec7} presents numerical examples and analyzes computational complexity. Finally, Section~\ref{conc} concludes the paper.

\section{Reliability problem}
\label{sec2}
In probabilistic reliability analysis, the performance of a system with a {$d_x$}-dimensional input variables $\mathbf{x} \subseteq \mathbb{R}^{d_x}$ is assessed using a limit state function (LSF)  $S: \mathbb{R}^{d_x} \rightarrow \mathbb{R}$ . This function divides the variable space \(\mathbb{R}^{d_x}\) into a safety domain and a failure domain, defined as follows:
\[
\begin{cases} 
S({x}) < 0, & \text{(failure)} \\
S({x}) = 0, & \text{(limit state)} \\
S({x}) > 0. & \text{(safe)}
\end{cases}
\]

We define {the probability density function (pdf)} $\pi(\mbf{x})$, which may represent the posterior given observed data $\mbf{y}$; however, we keep it more generic here. 
$\mathcal{F}=\{\mbf{x} \in \mathbb{R}^{d_x}: S(\mbf{x})\leq 0\}$ is the target failure domain. Let $\pi$ be the unnormalized target pdf of the random variable ${X}$, and $\widetilde{\pi}(\mathbf{x})=\frac{\pi(\mathbf{x})}{Z}$ is the normalized  pdf of the target under the availability of the normalizing constant $Z$. The corresponding failure probability  $P_f$ can be expressed as the evaluation of the integral over the failure space
\begin{equation}
\label{EQFAILURE}
    P_f \stackrel{\text{def.}}{=} \mathbb{P}\left(\mbf{X} \in \mathcal{F} \right) = \int \mathbb{I}_\mathcal{F}(\mbf{x})\widetilde{\pi}(\mbf{x}) dx,
\end{equation}
where $\mathbb{I}_{\mathcal{F}}(\mbf{x})=\mathbb{I}_{S(\mathbf{x}_k) \leq 0}$ is the failure indicator function that gives $1$ if $S(\mbf{x})<0$ and $0$ otherwise. {Equation} (\ref{EQFAILURE}) determines the failure probability, defined as the probability of the LSF being negative. When the dimension {$d_x$} is high (e.g., {$d_x > 20$)} or the failure boundary is complex (i.e., implicit or nonlinear), constructing and sampling from $\widetilde{\pi}(\mbf{x})$ becomes challenging. The main goal is to devise an efficient method for the approximation of the failure probability $P_f$ adequately.

\section{Preliminaries}
\label{sec3}
In this section, we briefly review the main concepts underlying FAMIS, namely adaptive importance sampling \cite{Bugallo7974876,ELVIRA201777} and normalizing flows (NFs) \cite{kobyzev2020normalizing}.

\subsection{Importance sampling}

Importance sampling (IS) is a variance reduction MC method that uses \textit{weighted} samples to perform inference with respect to the target distribution $\tilde{\pi}$ \cite{luengo2020survey}. It generates $K$ independent samples from an importance distribution, or proposal pdf, $q(\mbf{x})$, with $\mbf{x}_k \sim q(\mbf{x})$ for $k=1,...,K$. \blue{When $Z$ is known, the unnormalized IS (UIS) estimator is given by
\begin{equation}
    \widehat{I}_{\text{UIS}}=\frac{1}{K} \sum_{k=1}^{K} {w}_k \mathbb{I}_{S(\mathbf{x}_k) \leq 0},
\end{equation}
where $w_k=\frac{\tilde{\pi}(\mbf{x}_k)}{q(\mbf{x}_k)}$ denotes the importance weight assigned to each of the i.i.d. samples. When $Z$ is unknown, the self-normalized IS (SNIS) estimator \cite{elvira2021advances} can instead be employed:
\begin{equation}
   \mathbb{E}_{\pi} \left[ \mathbb{I}_{S(\mathbf{x}) \leq 0} \right] \approx \widehat{I}_{\text{SNIS}} =  \sum_{k=1}^{K} \bar{w}_k \mathbb{I}_{S(\mathbf{x}_k) \leq 0},
\end{equation}
where $\bar{w}_k= \frac{w_k}{\sum_{k=1}^{K} w_k}$ are the normalized weights, satisfying $\sum_{k=1}^{K} \bar{w}_{k}=1$.}

\subsection{Adaptive importance sampling (AIS)}

{For a general target $\pi(\mbf{x})$, the efficiency of $\widehat{I}_{\text{SNIS}}$ depends strongly on the proposal distribution $q(\mbf{x})$, motivating adaptive importance sampling (AIS) \cite{cornuet2012adaptive}. AIS methods iteratively refine one or more proposals over $t= 1, \dots, T$. In a generic setting, $K$ samples are generated from each of $N$ proposals, $\{q_n^{(t)}(\mbf{x}; \boldsymbol{\mu}_n^{(t)}, \boldsymbol{\Sigma}_n^{(t)})\}_{n=1}^{N}$, parametrized by the location $\boldsymbol{\mu}_n$ and covariance matrix $\mbf{\Sigma}_n$.
The resulting $NK$ samples at each iteration are assigned importance weights
$w_{n,k}^{(t)}= \widetilde{\pi}(\mbf{x}_{n,k}^{(t)})/q_{n}^{(t)}(\mbf{x}_{n,k}^{(t)})$. An important advance in AIS is deterministic mixture (DM) weighting, defined as
\begin{equation}
    w_{n,k}^{(t)} = \frac{\widetilde{\pi}(\mathbf{x}_{n,k}^{(t)})}{\Psi(\mathbf{x}_{n,k}^{(t)})},
\label{DMweights}
\end{equation}
where the equally weighted mixture proposal is denoted by \resizebox{0.33\linewidth}{!}{$\Psi(\mathbf{x}_{n,k}^{(t)}) \equiv \frac{1}{N} \sum_{n=1}^{N} q_n^{(t)}(\mathbf{x}_{n,k}^{(t)}; \boldsymbol{\mu}_n^{(t)}, \boldsymbol{\Sigma}_n^{(t)})$}. The location parameters $\{\boldsymbol{\mu}_n^{(t+1)}\}_{n=1}^N$ are subsequently updated at each iteration $t$.

Several strategies have been proposed for adapting the proposal family \cite{Bugallo7974876,ELVIRA201777}. A common approach relies on local or global resampling. The sampling, weighting, and resampling steps are repeated until a stopping criterion is satisfied, such as a maximum number of iterations $T$ \cite{Elvira8682284}. DM weighting is employed in deterministic mixture population Monte Carlo (DM-PMC) \cite{ELVIRA201777}, whose final-iteration estimator takes the form
\begin{equation}
\label{MISeqn}
    \widehat{P}_f =  \sum_{k=1}^{K} \sum_{n=1}^{N}  \bar{w}_{n,k}^{(T)} \mathbb{I}_{S(\mathbf{x}_{n,k}^{(T)}) \leq 0 } ,
\end{equation}}
where $\bar{w}_{n,k}^{(T)}$ denotes the normalized weights at the final iteration $T$. 


\subsection{Normalizing flows}
Normalizing flows (NFs) are a class of deep generative machine learning models designed for density estimation and sample generation \cite{kobyzev2020normalizing, rezende2015variational, papamakarios2021normalizing}. They provide a flexible framework to transform simple probability distributions into more complex ones through a 
sequence of invertible and differentiable mappings. 
The key idea is to begin with an easy-to-sample {base distribution}, 
denoted by $p_{0}(\mathbf{z})$, and apply a sequence of invertible 
transformations $\mathbf{x} = f_{\phi}(\mathbf{z})$, where the transformation $f_{\phi}$ is often referred to as the {generator}. 
This process maps samples from the base distribution into a more flexible 
{target}, or {induced}, distribution $p(\mathbf{x})$. 

Let $f_{\phi}(\mathbf{z}): \mathbb{R}^{d_x} \to \mathbb{R}^{d_x}$ denote a 
diffeomorphism parameterized by $\phi$. The 
change-of-variables formula establishes the density transformation
\begin{equation}
\label{changeofVar}
    p(\mathbf{x}) 
    = p_{0}(\mathbf{z}) 
      \left|\det \nabla_{\mathbf{z}} f_{\phi}(\mathbf{z})\right|^{-1},
\end{equation}
where $p$ denotes the pushforward of the base distribution 
$p_{0}$ under $f_{\phi}$, and $\det \nabla_{\mathbf{z}} f_{\phi}(\mathbf{z})$ 
is the Jacobian determinant of $f_{\phi}$. 
Thus, if $\mathbf{z}^{(i)} \sim p_{0}$, then 
$\mathbf{x}^{(i)} = f_{\phi}(\mathbf{z}^{(i)})$ are distributed 
according to $p(\mathbf{x})$. Let $\nabla_{\mathbf{z}} f_{\phi}(\mathbf{z})= J_{f_{\phi}}(\mathbf{z})$, Eq. (\ref{changeofVar}) can be equivalently rewritten as 
\begin{equation}
p(\mathbf{x})
\;=\;
p_{0}\big(f_{\phi}^{-1}(\mathbf{x})\big)\;
\Big|\det J_{f_{\phi}^{-1}}(\mathbf{x})\Big|.
\label{eq:flow-density}
\end{equation}
The inverse transformation $f_{\phi}^{-1}$ maps samples from the induced 
distribution back to the base distribution, which motivates the name 
\emph{normalizing flows}, since the multivariate standard normal is 
frequently chosen as the base density. 
However, any distribution from which efficient sampling is possible 
can serve as a base.

To model complex target distributions in high-dimensional spaces, 
a single transformation is often insufficient. 
Instead, NFs compose multiple invertible mappings:
\begin{equation}
    f_{\phi} = f_{\phi_L,L} \circ f_{\phi_{L-1},L-1} \circ \cdots \circ f_{\phi_1,1},
    \label{NF_trans_eqn}
\end{equation}
where $L$ is the number of flow layers \cite{kobyzev2020normalizing}. Since the composition of 
bijective functions is also bijective, the resulting transformation is 
invertible. The Jacobian determinant factorizes in the log domain as
\begin{equation}
\label{logeqn}
    \log \left|\operatorname{det} J_{f_{\phi}}(\mathbf{z})\right|
    =
    \sum_{\ell=1}^{L} 
    \log \left|\operatorname{det} J_{f_{\phi_\ell,\ell}}\left(\mathbf{h}_{\ell-1}\right)\right|,
\end{equation}
with $\mathbf{h}_0=\mathbf{z}$ and
$\mathbf{h}_{\ell}=f_{\phi_\ell,\ell}(\mathbf{h}_{\ell-1})$,
$\ell=1,\ldots,L$.
Equation~(\ref{logeqn}) represents a layerwise log-determinant decomposition. 

The design of each flow layer ensures both invertibility and efficient 
Jacobian determinant computation, which are critical for practical 
training and inference. Popular flow architectures include planar flows, 
radial flows, affine coupling layers, and autoregressive flows \cite{rezende2015variational,dinh2014nice,dinh2016realnvp,kingma2016iaf,papamakarios2017maf,durkan2019neural}. 
These architectures vary in expressive power and computational cost, but all transform tractable base distributions into expressive induced distributions suitable for tasks 
such as importance sampling and rare-event 
probability computation.

\section{The proposed FAMIS algorithm}
\label{sec4}

In this section, we present the repulsive \textit{flow-based adaptive multiple importance sampling} (FAMIS) framework. The core idea is to leverage  normalizing flows (NFs) within a multiple importance sampling (MIS) framework, while ensuring that the components of the mixture explore the failure space efficiently through a Jensen-Shannon divergence (JSD) repulsion term.
The proposed method constructs an adaptive IS mixture whose components are obtained by transforming simple Gaussian base densities through a shared NF. The algorithm takes as an input a flow model, $N$ proposals, $K$ samples over $T$ iterations and outputs an estimate of failure probability $P_f$. Failure events are generated after learning the proposal distributions by generating samples in the latent space and mapping them back to target space.

{Algorithm~\ref{alg:famis} summarizes the FAMIS workflow. {Given target density $\tilde{\pi}$}, the algorithm proceeds through \(T\) adaptive iterations and consists of four main stages. In Step~1, the flow-induced proposal components \(\{q_n^{(t)}(\mathbf{x};\boldsymbol{\theta}_n,{\phi})\}_{n=1}^N\) are constructed through a shared normalizing flow, parameterized by $\phi$, and combined into the weighted mixture \(q_\alpha^{(t)}\). This mixture is then combined with the defensive density \(\pi_0\) to form the sampling density \({\Psi}^{(t)}\). 
In Step~2, \(\{\mathbf{x}_k^{(t)}\}_{k=1}^K\) samples from \({\Psi}^{(t)}\) are used to evaluate the LSF \(S(\mathbf{x}_k^{(t)})\), and the smooth rare-event surrogate \(s^{(t)}(\mathbf{x})\) is used to define the tempered target \(\pi_\star^{(t)}\).  Step~3 adapts the proposal by updating the mixture weights and the flow and component parameters. The detailed proposal adaptation rules, including the training objective, are detailed in Section~\ref{sec5}. Finally, Step~4 draws samples from the learned mixture \(q_\alpha^{(T)}\) and computes the failure probability using a deterministic-mixture estimator.}

\begin{algorithm}[!htbp]
\caption{FAMIS}
\label{alg:famis}
\begin{algorithmic}[1]
\REQUIRE Target density $\widetilde{\pi}(\mathbf{x})$, LSF $S(\mathbf{x})$, number of proposals $N$, {training batch size $K$}, iterations $T$, schedules $\{\kappa^{(t)}\}$, $\{u^{(t)}\}$, $\{\gamma^{(t)}\}$, repulsion weight $\lambda_{\mathrm{JSD}}$.

\STATE 
Initialize shared flow parameters ${\phi}^{(1)}$, component parameters $\{\boldsymbol{\theta}_n^{(1)}\}_{n=1}^N$, and mixture weights $\{\alpha_n^{(1)}\}_{n=1}^N$.

\FOR{$t=1,\ldots,{T}$}
    \Statex {\textbf{Step 1: Flow mixture and defensive sampling.}}
    \Statex 
    \begin{enumerate}[label=(\alph*), leftmargin=2em, labelwidth=1em, align=left]
    \item Construct the flow proposals
    $q_n^{(t)}(\mathbf{x};\boldsymbol{\theta}_n,{\phi})$ using Eqs.~{(\ref{pushF_eqn})--(\ref{eq:flow_logdensity})} and the adaptive mixture using Eq. (\ref{eq10})
    \vspace{-0.55 cm}
    \[
    q_{\alpha}^{(t)}(\mathbf{x})
    =
    \sum_{n=1}^{N}\alpha_n^{(t)}
    q_n^{(t)}(\mathbf{x};\boldsymbol{\theta}_n,{\phi}).
    \]
    \item Form the defensive exploration mixture using a defensive density $\pi_0$
    \[
    {\Psi}^{(t)}(\mathbf{x})
    =
    \bigl(1-\epsilon^{(t)}\bigr)q_{\alpha}^{(t)}(\mathbf{x})
    +
    \epsilon^{(t)}\pi_0(\mathbf{x}), \qquad 0<\epsilon^{(t)}<1.
    \]

    \item Draw $K$ samples
   $ \{\mathbf{x}_k^{(t)}\}_{k=1}^{K}\sim {\Psi}^{(t)}(\mathbf{x})$.
    \end{enumerate}
    
    \Statex {\textbf{Step 2: Rare-event surrogate and annealing.}}
    \Statex 
    \begin{enumerate}[label=(\alph*), leftmargin=2em, labelwidth=1em, align=left]
    \item Compute the rare-event surrogate $s^{(t)}(\mathbf{x}_k^{(t)})$ and the tempered surrogate target
    $\pi_\star^{(t)}$ using Eqs.~(\ref{eq:log_rare_event_surrogate_target})--(\ref{gamma_update}).

    \item Compute the posterior responsibilities 
    $r_n^{(t)}(\mathbf{x}_k^{(t)})
    =
    \frac{
    \alpha_n^{(t)}q_n^{(t)}(\mathbf{x}_k^{(t)})
    }{
    q_{\alpha}^{(t)}(\mathbf{x}_k^{(t)})
    }$, for $n=1,\ldots,N.
    $
    \end{enumerate}
    \Statex {\textbf{Step 3: Mixture-weight and flow-parameter update.}}
    \Statex  
    \begin{enumerate}[label=(\alph*), leftmargin=2em, labelwidth=1em, align=left]
    \item Compute the normalized surrogate weights
    $\{\bar{w}_k^{(t)}\}_{k=1}^{K}$ using Eq.~(\ref{eq:normalized_forward_weights}).

    \item  Update the mixture weights $\alpha_n$ by the Rao--Blackwellized rule
    \[
    \alpha_n^{(t+1)}
    =
    \sum_{k=1}^{K}
    \bar{w}_k^{(t)}
    r_n^{(t)}(\mathbf{x}_k^{(t)}),
    \qquad n=1,\ldots,N.
    \]
\vspace{-0.2 cm}
    \item Compute the hybrid fit loss $\mathcal{L}_{\mathrm{fit}}^{(t)}$ using Eq.~(\ref{eq:hybrid_kl_fit}) and the JSD repulsion term $\mathcal{R}_{\mathrm{JSD}}^{(t)}$ using Eq.~(\ref{eq:jsd_defensive_estimator}).

    \item  Update the flow and component parameters using learning rate
$\eta$ by minimizing
\vspace{-0.3 cm}
    \[
    \mathcal{L}^{(t)}(\boldsymbol{\theta},{\phi})
    =
    \mathcal{L}_{\mathrm{fit}}^{(t)}
    -
    \lambda_{\mathrm{JSD}}\mathcal{R}_{\mathrm{JSD}}^{(t)}.
    \]
    \[
    (\boldsymbol{\theta},{\phi})
    \leftarrow
    (\boldsymbol{\theta},{\phi})
    -
    \eta\nabla_{\boldsymbol{\theta},{\phi}}
    \mathcal{L}^{(t)} .
    \]
    \end{enumerate}
\ENDFOR

\Statex {\textbf{Step 4:  Failure estimation.}}
\Statex 
\begin{enumerate}[label=(\alph*), leftmargin=2em, labelwidth=1em, align=left]
\item Draw
$\{\mbf{x}_k^{({T})}\}_{k=1}^{{K_{\mathrm{est}}}}
 \overset{\mathrm{i.i.d.}}{\sim}q_{\alpha}^{({T})}$
and compute $ w_k^{({T})}
    =\frac{\widetilde{\pi}(\mbf{x}_k^{({T})})}
           {q_{\alpha}^{({T})}(\mbf{x}_k^{({T})})}$, {where \(K_{\mathrm{est}}\) is the estimation sample size}.

\item Return the deterministic-mixture IS estimate
\vspace{-0.3 cm}
\[
\widehat{P}_f
=
\frac{1}{{K_{\mathrm{est}}}}
\sum_{k=1}^{{K_{\mathrm{est}}}}
w_k^{(T)} \mathbb{I}_{S(\mathbf{x}_k^{(T)})\leq0}.
\]
\end{enumerate}
\end{algorithmic}
\end{algorithm}

{\subsection{Flow mixture proposal and defensive sampling (Step 1)}}
{The first step defines the sampling family used by FAMIS. We construct a mixture of flow proposal components and augment it during training with a broad defensive distribution to explore failure regions not yet covered by the learned mixture.}

\noindent\textbf{Flow composition and induced proposals.} 
We learn $N$ proposal distributions $\{q_n(\mathbf{x};\boldsymbol{\theta}_n,\phi)\}_{n=1}^{N}$ and their mixture 
\begin{equation}
    q_{{\alpha}}(\mathbf{x}) = \sum_{n=1}^N \alpha_n q_n(\mathbf{x};\boldsymbol{\theta}_n,\phi),
    \label{eq10}
\end{equation} 
where $\alpha_n \ge 0$ are the mixture weights such that $\sum_{n=1}^{N} \alpha_n = 1$, and $\boldsymbol{\theta}_n=(\boldsymbol{\mu}_n, \boldsymbol{\Sigma}_n)$ are the adaptive parameters. Each component is obtained by pushing a Gaussian base
$q_{0,n}=\mathcal{N}(\boldsymbol{\mu}_n,\boldsymbol{\Sigma}_n)$ through a shared invertible transformation (i.e., RealNVP flow)
$f_\phi:\mathbb{R}^{d_x}\to\mathbb{R}^{d_x}$,
with $\phi=(\phi_1,\ldots,\phi_L)$, defined as the composition of
$L$ invertible and trainable transformations
$\{f_{\phi_\ell,\ell}\}_{\ell=1}^{L}$.
Specifically, $f_{\phi}$
is implemented as affine coupling layers as in {Eq.}~(\ref{NF_trans_eqn}), so that $\mathbf{x}=f_{\phi}(\mathbf{z})$,
where $\mathbf{z} \sim q_{0,n}$ and $\mathbf{z}=f_{{\phi}}^{-1}(\mathbf{x})$ is the inverse mapping.
Therefore, the corresponding pushforward proposal density $q_n$ is given by the change-of-variables formula in the original space as
\begin{equation}
    q_n(\mathbf{x};\boldsymbol{\theta}_n,\phi)
    =
    q_{0,n}\left(f_{\phi}^{-1}(\mathbf{x})\right)
    \left|
        \det \mathbf{J}_{f_{\phi}^{-1}}(\mathbf{x})
    \right|,
    \label{pushF_eqn}
\end{equation}
where
$\mathbf{J}_{f_\phi^{-1}}(\mathbf{x})
=\nabla_{\mathbf{x}}f_\phi^{-1}(\mathbf{x})
=\frac{\partial f_\phi^{-1}(\mathbf{x})}{\partial\mathbf{x}}$. Equation (\ref{pushF_eqn}) can be rewritten as
\begin{equation}
\begin{aligned}
\log q_n(\mathbf{x};\boldsymbol{\theta}_n,\phi)
&=
\log q_{0,n}(\mathbf{z})
+
\log\left|\det J_{f_\phi^{-1}}(\mathbf{x})\right|\\
&=
\log q_{0,n}(\mathbf{z})
+
\sum_{\ell=1}^{L}
\log\left|
\det J_{f_{\phi_\ell,\ell}^{-1}}(\mathbf{h}_{\ell})
\right|\\
&=
\log q_{0,n}(\mathbf{z})
-
\sum_{\ell=1}^{L}
\log\left|
\det J_{f_{\phi_\ell,\ell}}(\mathbf{h}_{\ell-1})
\right|,
\end{aligned}
\label{eq:flow_logdensity}
\end{equation}
where $\mathbf{z}=f_\phi^{-1}(\mathbf{x})$,
$\mathbf{h}_{L}=\mathbf{x}$, and
$\mathbf{h}_{\ell-1}
=f_{\phi_\ell,\ell}^{-1}(\mathbf{h}_{\ell})$,
for $\ell=L,\ldots,1$. The final equality in Eq. (\ref{eq:flow_logdensity}) follows from the inverse-Jacobian identity $J_{f_{\phi_\ell,\ell}^{-1}}(\mathbf{h}_{\ell})
=
[
J_{f_{\phi_\ell,\ell}}(\mathbf{h}_{\ell-1})
]^{-1}$ and Eq. (\ref{logeqn}).
 By sharing \({\phi}\) across $\{q_n\}_{n=1}^N$, FAMIS learns a shared nonlinear transformation that captures the global geometry of the target $\widetilde{\pi}$, while the base parameters \(\{ \boldsymbol{\mu}_n,\boldsymbol{\Sigma}_n\}_{n=1}^N\) of $q_n$ capture its different local modes.

\noindent\textbf{{Defensive exploration mixture.}} With a shared flow, all mixture components are coupled through the same transport map. While this improves efficiency, it can also make the adaptation sensitive to early mode selection: regions not explored by the current mixture may receive little or no gradient information. To mitigate this effect, we introduce a defensive exploration mixture \cite{hesterberg1995weighted,cappe2008adaptive}. 
During training, the learned mixture $q_{{\alpha}}$ is subsequently combined with a defensive exploration distribution
$\pi_0$, and samples are drawn from the mixture
\begin{equation}
    {\Psi}^{(t)}(\mathbf{x})
    =
    (1-\epsilon^{(t)})q_{{\alpha}}(\mathbf{x})
    +
    \epsilon^{(t)} \pi_0(\mathbf{x}),
\label{eq:defensive_mixture}
\end{equation}
where $0<\epsilon^{(t)}<1$ is the mixing coefficient that controls the proportion of defensive samples. 
The defensive distribution $\pi_0$ is chosen to be broad to provide exploratory samples that can reveal tail regions not yet covered by $q_{{\alpha}}$.  We consider a Cauchy distribution for its heavy tails that provide broad exploration during adaptation and increase the chance of sampling failure regions poorly covered by the proposal:
\begin{equation}
    \pi_0(\mathbf{x})
    =
    \prod_{j=1}^{d_x}
    \frac{1}{\pi s_c\left(1+(x_j/s_c)^2\right)},
    \label{eq:cauchy_defensive}
\end{equation}
where $s_c>0$ is a scale parameter controlling the exploratory spread. Given samples $\{\mathbf{x}_k\}_{k=1}^K\sim {\Psi}^{(t)}$, the corresponding log-density, $\log {\Psi}^{(t)}(\mathbf{x})$, is available in closed form and is used to evaluate the sampling density ${\Psi}^{(t)}$ in \eqref{eq:defensive_mixture}.

{\subsection{Rare-event surrogate and annealing (Step 2)}}
\label{subsec:defensive_mixture}

\begin{figure}[h]
\centering 
\includegraphics[width=0.45\textwidth]{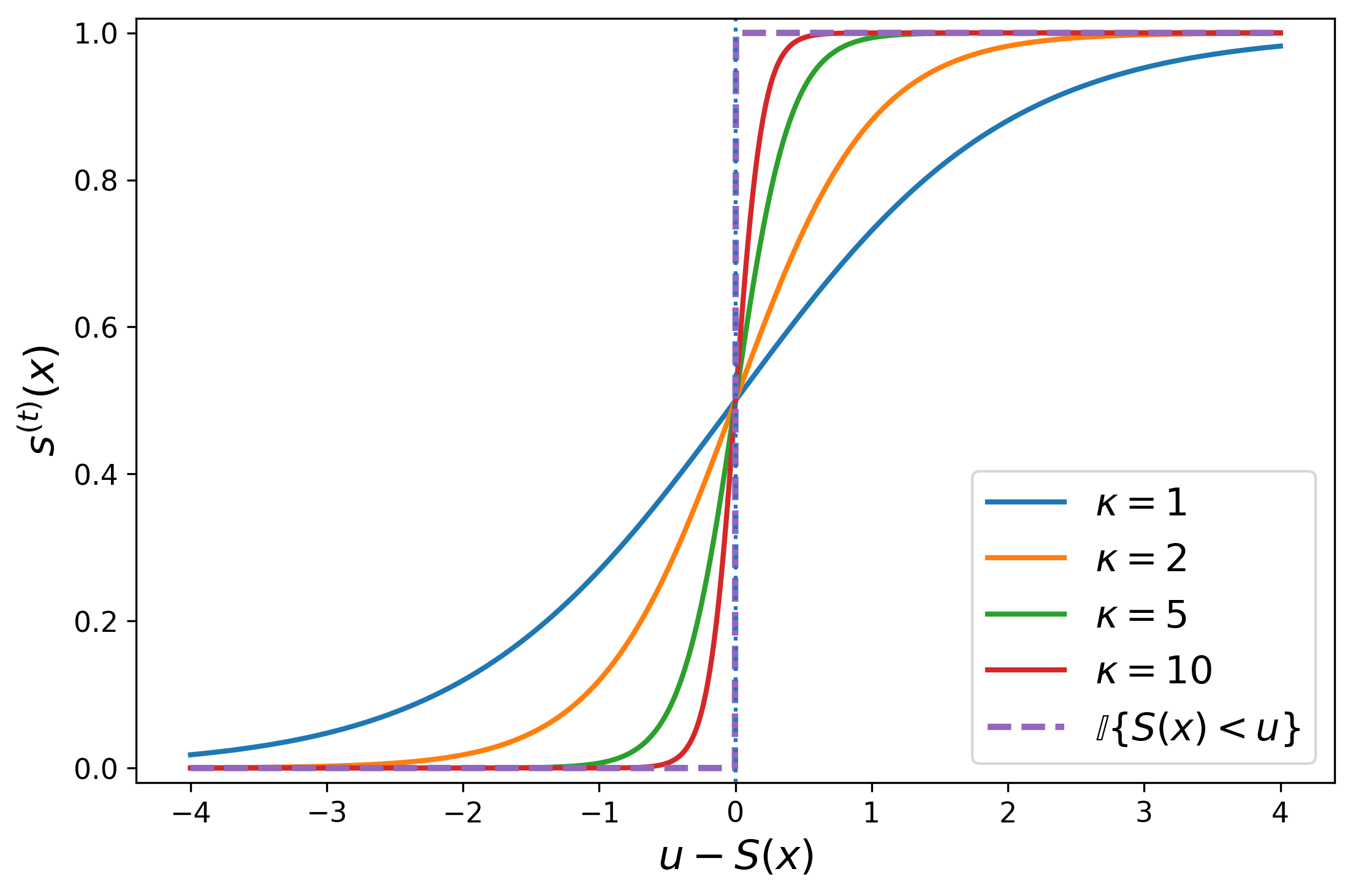}
\caption{Convergence of the smooth rare-event surrogate as $\kappa$ increases. The surrogate is
$s(\mathbf{x})=\big(1+\exp(\kappa(S(\mathbf{x})-u))\big)^{-1}$ and is plotted
as a function of $u-S(\mathbf{x})$ for several values of $\kappa$ (with $u=0$). }
\label{surrogate_alpha}
\end{figure}

The hard (i.e., discontinuous) indicator $\mathbb{I}_{S(\mathbf{x})\le 0}$ is non-differentiable and provides little to no gradient information when failures are rare in early iterations. To guide the {defensive mixture ${\Psi}^{(t)}$} toward the quasi-optimal IS density $\widetilde{\pi}(\mathbf{x}\mid \mathcal{F})$ without requiring presampled failures, we therefore introduce a smooth
sigmoid surrogate $s^{(t)}(\mathbf{x}) \in(0,1)$,
\begin{equation}
s^{(t)}(\mathbf{x})
\begin{aligned}[t]
&\;\triangleq\;
\sigma\!\left[-\kappa^{(t)}\{S(\mathbf{x})-u^{(t)}\}\right] \\
&\;=\;\Big(1+\exp\!\big(\kappa^{(t)}(S(\mathbf{x})-u^{(t)})\big)\Big)^{-1}
\end{aligned},
\label{eq:sigmoid}
\end{equation}
where \(\sigma(\cdot)\) is the logistic sigmoid, $\kappa^{(t)}> 0$ is an increasing
sharpness parameter, and $u^{(t)}$ is a decreasing shift parameter.
As $u^{(t)}\to 0$ and $\kappa^{(t)}$ increases, the surrogate concentrates on the true failure set $\mathcal{F}=\{\mathbf{x}:S(\mathbf{x})\le 0\}$.
The surrogate (\ref{eq:sigmoid}) satisfies $\lim_{\kappa^{(t)}\to\infty}
    s^{(t)}(\mathbf{x})
    =
    \mathbb{I}_{S(\mathbf{x})<u^{(t)}}$, 
 $S(\mathbf{x})\neq u^{(t)}$.
Figure \ref{surrogate_alpha} illustrates
the convergence of the approximation.
The corresponding unnormalized rare-event surrogate target is
\begin{equation}
    {\pi}^{(t)}(\mathbf{x})
    =
    \widetilde{\pi}(\mathbf{x})s^{(t)}(\mathbf{x}),
    \label{eq:rare_event_surrogate_target}
\end{equation}
or equivalently,
\begin{equation}
    \log {\pi}^{(t)}(\mathbf{x})
    =
    \log \widetilde{\pi}(\mathbf{x})+\log s^{(t)}(\mathbf{x}).
    \label{eq:log_rare_event_surrogate_target}
\end{equation}
To avoid an overly concentrated target during the early iterations, we use a tempered version of the surrogate target,
\begin{equation}
    \pi^{(t)}_{\star}(\mathbf{x})
    \propto
    \left[
        {\pi}^{(t)}(\mathbf{x})
    \right]^{1/\gamma^{(t)}}
    =
    \left[
        \widetilde{\pi}(\mathbf{x})s^{(t)}(\mathbf{x})
    \right]^{1/\gamma^{(t)}},
    \label{eq:tempered_rare_event_target}
\end{equation}
where \(\gamma^{(t)}\geq 1\) is a temperature parameter that decreases to one, and $\pi(\mathbf{x})$ denotes the unnormalized target density. 
During training, $\gamma^{(t)}$ is updated as
  \begin{equation}
  \label{gamma_update}
      \gamma^{(t)} = \max \left\{ \gamma_{\min}, \gamma_{\text{init}} - \left( \frac{\gamma_{\text{init}} - \gamma_{\min}}{T^{'}} \right)  t\right\},
  \end{equation}
where $\gamma_{\mathrm{init}}$ is the initial temperature, $\gamma_{\min}\ge1$ is the final temperature, $T'$ is the number of annealing iterations, and $t$ is the current iteration. Large values of \(\gamma^{(t)}\) flatten the target during the early iterations, while \(\gamma^{(t)}=1\) recovers the final rare-event surrogate.

\section{Proposal adaptation}
\label{sec5}
{This section details {Steps~3 and 4 }of Algorithm~\ref{alg:famis}. After the defensive samples and the tempered surrogate have been constructed, FAMIS adapts the proposal in two complementary ways. 
First, the normalized surrogate weights are computed and the mixture weights are updated via Rao--Blackwellized responsibilities (Steps~3a--b). Second, the shared flow and component parameters are updated by minimizing a repulsive training objective (Steps~3c--d).
{Finally, FAMIS uses the final learned mixture \(q_\alpha^{(T)}\) as the importance sampling density to compute the failure probability (Step~4).}

\subsection{{Adaptive update of mixture weights (Step 3a--b)}}
The mixture weights are updated separately from the gradient step on $(\boldsymbol{\theta},\phi)$. This follows the Rao--Blackwellized mixture adaptation principle \cite{cappe2008adaptive, douc2008convergence}, where component indicators are replaced by their posterior
responsibilities $r_n^{(t)}(\mathbf{x}_k)$ to reduce variance.
 For any $\mathbf{x}$, define the posterior responsibility, i.e., soft weights, of component \(n\) as
\begin{equation}
    r_n(\mathbf{x})
    \triangleq
    \frac{\alpha_n q_n(\mathbf{x};\boldsymbol{\theta}_n,\phi)}
    {q_{\alpha}(\mathbf{x})}.
\label{eq:resp}
\end{equation}
Given a shared batch \(\{\mathbf{x}_k\}_{k=1}^{K}\sim {\Psi}^{(t)}\), define the unnormalized
importance weights

\begin{equation}
     {w}^{(t)}_k
    = \frac{\pi^{(t)}_{\star}(\mathbf{x}_k)}{{\Psi}^{(t)}(\mathbf{x}_k)} =
    \frac{
       \left(\widetilde{\pi}(\mathbf{x}_k) s^{(t)}(\mathbf{x}_k)
\right)^{1/\gamma^{(t)}}}{ {\Psi}^{(t)}(\mathbf{x}_k)}.
\label{weight_eqn}
\end{equation}
Equivalently, writing $\log {\pi}^{(t)}(\mathbf{x})
    =
    \log \widetilde{\pi}(\mathbf{x})+\log s^{(t)}(\mathbf{x})$,
we compute the normalized weights as
    \begin{equation}
    \bar{w}_k^{(t)}
    =
    \frac{
    \exp\left[
        \frac{1}{\gamma^{(t)}}
        \log {\pi}^{(t)}(\mathbf{x}_k)
        -
        \log {\Psi}^{(t)}(\mathbf{x}_k)
    \right]
    }{
    \sum_{k=1}^{K}
    \exp\left[
        \frac{1}{\gamma^{(t)}}
        \log {\pi}^{(t)}(\mathbf{x}_k)
        -
        \log {\Psi}^{(t)}(\mathbf{x}_k)
    \right]
    }.
    \label{eq:normalized_forward_weights}
\end{equation}
The updated mixture weights are then obtained using a Rao--Blackwellized responsibility update
\begin{equation}
   {\alpha}_{n}^{(t+1)}
    =
    \sum_{k=1}^{K}
    \bar{w}^{(t)}_k r^{(t)}_n(\mathbf{x}_k). 
    \label{alpha_weights}
\end{equation}
The update is analogous to the mixture weight adaptation used
in population Monte Carlo methods \cite{ELVIRA201777,Bugallo7974876}.

\subsection{{Parameter update and repulsive training objective (Step 3c--d)}}
At iteration \(t\), samples are drawn from the defensive mixture $\{\mathbf{x}_k^{(t)}\}_{k=1}^{K} \sim {\Psi}^{(t)}(\mathbf{x})$.
For each sample, all component log-densities $\log q_n(\mathbf{x}_k;\boldsymbol{\theta}_{n}^{(t)},\phi^{(t)}), n=1,\ldots,N$,
are evaluated using a single inverse pass through the shared flow $f_{\phi}$. Following the posterior responsibility $r_n(\mathbf{x})$ in Eq. \eqref{eq:resp}, then,
\begin{equation}  
\frac{q_n(\mathbf{x};\boldsymbol{\theta}_n,\phi)}
    {q_{\alpha}(\mathbf{x})}
    =
    \frac{r_n(\mathbf{x})}{\alpha_n}.
\label{identity_1}
\end{equation}
Therefore, for any integrable function \(h\),

\begin{equation}
    \mathbb{E}_{q_{{\alpha}}}[h(\mathbf{x})]
    =
    \mathbb{E}_{{\Psi}^{(t)}}
    \left[
        \frac{q_{{\alpha}}(\mathbf{x})}{{\Psi}^{(t)}(\mathbf{x})}
        h(\mathbf{x})
    \right].
    \label{eq:defensive_expectation_identity}
\end{equation}

Similarly, component-wise expectations can be estimated from the same defensive samples by 
\begin{equation}
    \mathbb{E}_{q_n}[h(\mathbf{x})]
    =
    \mathbb{E}_{{\Psi}^{(t)}}
    \left[
        \frac{q_{{\alpha}}(\mathbf{x})}{{\Psi}^{(t)}(\mathbf{x})}
        \frac{r_n(\mathbf{x})}{\alpha_n}
        h(\mathbf{x})
    \right].
    \label{identity_2}
\end{equation}
For uniform mixture weights, $\alpha_n=1/N$, this reduces to
    $\mathbb{E}_{q_n}[h(\mathbf{x})]
    =
    N\,
    \mathbb{E}_{{\Psi}^{(t)}}
    \left[
        \frac{q_{{\alpha}}(\mathbf{x})}{{\Psi}^{(t)}(\mathbf{x})}
        r_n(\mathbf{x})h(\mathbf{x})
    \right]$.
The flow and mixture parameters are then optimized by minimizing an objective loss function composed of a
hybrid fit term and a repulsive regularization term. The adaptation objective consists of two terms,  as described in the next subsection.

\noindent\textbf{{Hybrid Kullback–Leibler divergence fit term.}}
The fit term is a hybrid Kullback–Leibler (KL) divergence that combines a reverse-KL (RKL) refinement term and a forward-KL (FKL) discovery term, i.e., cross entropy.

The reverse-KL term fits the aggregate proposal \(q_{{\alpha}}\) toward the tempered surrogate target
\(\widetilde{\pi}^{(t)}\):
\begin{equation}
    \mathrm{KL}
    \left(
        q_{{\alpha}}
        \,\middle\|\,
        \pi_{\star}^{(t)}
    \right)
    =
    \mathbb{E}_{q_{\boldsymbol{\alpha}}}
    \left[
        \log q_{{\alpha}}(\mathbf{x})
        -
        \log \pi_{\star}^{(t)}(\mathbf{x})
    \right].
    \label{eq:rkl_exact}
\end{equation}
Using the identity \eqref{eq:defensive_expectation_identity}, Eq. \eqref{eq:rkl_exact} is estimated from defensive samples $\mathbf{x}_k \sim \Psi^{(t)}$ as
\begin{equation}
    \mathcal{L}_{\mathrm{RKL}}^{(t)}
    =
    \mathbb{E}_{{\Psi}^{(t)}}
    \left[
        \frac{
            q_{{\alpha}}(\mathbf{x})
        }{
            {\Psi}^{(t)}(\mathbf{x})
        }
        \left(
            \log q_{{\alpha}}(\mathbf{x})
            -
            \frac{1}{\gamma^{(t)}}
            \log \widetilde{\pi}^{(t)}(\mathbf{x})
        \right)
    \right].
    \label{eq:rkl_defensive}
\end{equation}
{Equation} \eqref{eq:rkl_defensive} adapts the current proposal in regions where \(q_{{\alpha}}\) already assigns probability mass.

 To improve mode discovery, we include the forward-KL term 
 \begin{equation}
    \mathrm{KL}
    \left(
        \pi_{\star}^{(t)}
        \,\middle\|\,
        q_{{\alpha}}
    \right)
    =
    \mathbb{E}_{\pi_{\star}^{(t)}}
    \left[
        \log \pi_{\star}^{(t)}(\mathbf{x})
        -
        \log q_{{\alpha}}(\mathbf{x})
    \right].
    \label{eq:fkl_exact}
\end{equation}
 The first term does not depend on the proposal parameters $\boldsymbol{\theta}$, so minimizing \eqref{eq:fkl_exact} is equivalent to minimizing
    $-\mathbb{E}_{\pi_{\star}^{(t)}}
    \left[
        \log q_{{\alpha}}(\mathbf{X})
    \right]$. Since direct sampling from \(\pi_{\star}^{(t)}\) is not available, Eq. \eqref{eq:fkl_exact} is estimated under the defensive mixture \({\Psi}^{(t)}\), using normalized weights \eqref{eq:normalized_forward_weights}.
  The forward KL term is then
\begin{equation}
    \mathcal{L}_{\mathrm{FKL}}^{(t)}
    =
    -
    \sum_{k=1}^{K}
    \bar{w}_k^{(t)}
    \log q_{{\alpha}}(\mathbf{x}_k).
    \label{eq:fkl_discovery}
\end{equation}
The hybrid fit objective is finally written as
\begin{equation}
    \mathcal{L}_{\mathrm{fit}}^{(t)}
    =
    \left(1-\beta^{(t)}\right)
    \mathcal{L}_{\mathrm{RKL}}^{(t)}
    +
    \beta^{(t)}
    \mathcal{L}_{\mathrm{FKL}}^{(t)},
    \label{eq:hybrid_kl_fit}
\end{equation}
where \(\beta^{(t)}\in[0,1]\) controls the trade-off between reverse-KL and forward-KL terms (i.e., larger values of \(\beta^{(t)}\) encourage exploration and mode discovery).

\noindent\textbf{{Repulsion via Jensen--Shannon divergence.}}
To discourage component overlap and promote coverage of multiple failure modes, the training objective involves a repulsive
regularizer based on the Jensen--Shannon divergence (JSD) between each component $q_n$ and the mixture $q_{\alpha}$. The corresponding JSD is
$\mathrm{JSD}(q_n\|q_\alpha)\triangleq\frac12\mathrm{KL}(q_n\|M_n)+\frac12\mathrm{KL}(q_\alpha\|M_n)$,
where $M_n\;\triangleq\;\big(q_n+q_\alpha\big)/2$. The repulsion term is then $\sum_{n=1}^{N}
    \mathrm{JSD}
    \left(
        q_n
        \,\middle\|\,
        q_{{\alpha}}
    \right)$, or equivalently,
\begin{align}
    \mathcal{R}_{\mathrm{JSD}}^{(t)}
    =
    \frac{1}{2}
    \sum_{n=1}^{N}
    \Bigg[
    &
    \mathbb{E}_{q_n^{(t)}}
    \left[
        \log
        \frac{q_n^{(t)}(\mathbf{x};\boldsymbol{\theta}_n,\phi)}{M_n^{(t)}(\mathbf{x})}
    \right]
    \nonumber\\
    &
    +
    \mathbb{E}_{q_{{\alpha}}^{(t)}}
    \left[
        \log
        \frac{q_{{\alpha}}^{(t)}(\mathbf{x})}{M_n^{(t)}(\mathbf{x})}
    \right]
    \Bigg],
\label{JSD_B4}
\end{align}
with $M_n^{(t)}=(q_n^{(t)}+q_\alpha^{(t)})/2$. Using Eqs. \eqref{eq:defensive_expectation_identity} and \eqref{identity_2}, the estimate of the JSD repulsion can be rewritten under ${\Psi}^{(t)}$ as 
\begin{equation}
\begin{aligned}
    \mathcal{R}_{\mathrm{JSD}}^{(t)}
    =
    \frac{1}{2}
    \sum_{n=1}^{N}
    \mathbb{E}_{{\Psi}^{(t)}}
    \Bigg[
    &
    \frac{q_n^{(t)}(\mathbf{x})}
{\Psi^{(t)}(\mathbf{x})}
    \left[ \log
        \frac{q_n^{(t)}(\mathbf{x};\boldsymbol{\theta}_n,\phi)}{M_n^{(t)}(\mathbf{x})}
    \right]
    \\
    &+
    \frac{q_{{\alpha}}^{(t)}(\mathbf{x})}{{\Psi}^{(t)}(\mathbf{x})}
    \left[ \log
        \frac{q_{{\alpha}}^{(t)}(\mathbf{x})}{M_n^{(t)}(\mathbf{x})}
    \right]
    \Bigg].
\end{aligned}
\label{eq:jsd_defensive_estimator}
\end{equation}
Both expectations in \eqref{eq:jsd_defensive_estimator} are hence estimated using the same batch sampled from \({\Psi}^{(t)}\). 
The overall objective minimized at iteration $t$ is then

\begin{equation}
    \mathcal{L}^{(t)}(\boldsymbol{\theta},\phi)
    =
    \mathcal{L}_{\mathrm{fit}}^{(t)}
    -
    \lambda_{\mathrm{JSD}}
    \mathcal{R}_{\mathrm{JSD}}^{(t)},
    \label{eq:full_objective_with_jsd}
\end{equation}
where \(\lambda_{\mathrm{JSD}}>0\) controls the strength of the repulsion.

\subsection{{Final importance sampling estimator (Step 4)}} 
After \(T\) adaptation iterations, FAMIS returns the final mixture proposal
\begin{equation}
    q_{{\alpha}}^{(T)}(\mathbf{x})
    =
    \sum_{n=1}^{N}
    \alpha_n^{(T)} q_n^{(T)}(\mathbf{x};\boldsymbol{\theta}_n,{\phi}).
\end{equation}
The failure probability is then estimated using \(q_{{\alpha}}^{(T)}(\mathbf{x})\) as the final importance sampling density. The computation can be summarized as follows.

\textit{Step 1:} Train the adaptive mixture of flow-based proposals for \(T\) iterations. At each iteration, minimize the loss function Eq.~(\ref{eq:full_objective_with_jsd}).

\textit{Step 2:} Estimate \(P_f\) using importance sampling with \(q_{{\alpha}}^{(T)}(\mathbf{x})\) as the final sampling density. {Let
\(K_{\mathrm{est}}\) denote the number of samples used only for the final
failure probability estimation}:
\begin{enumerate}
    \item Generate \(K_{\mathrm{est}}\) i.i.d. samples
    \[
    \mathbf{x}_{k}^{(T)}\sim q_{{\alpha}}^{(T)}(\mathbf{x}), 
    \qquad k=1,\ldots,K_{\mathrm{est}} .
    \]
    
    \item Evaluate the LSF \(S(\mathbf{x}_{k}^{(T)})\) for each generated sample.
    
    \item Compute the importance weights
    \[
    w_k^{(T)}
    =
    \frac{\widetilde{\pi}(\mathbf{x}_{k}^{(T)})}
    {q_{{\alpha}}^{(T)}(\mathbf{x}_{k}^{(T)})}.
    \]
    
    \item Estimate the failure probability as
    \begin{equation}
        \widehat{P}_{{f}}
        =
        \frac{1}{K_{\mathrm{est}}}
        \sum_{k=1}^{K_{\mathrm{est}}}
        w_{k}^{(T)}
        \mathbb{I}_{S(\mathbf{x}_{k}^{(T)})\leq 0}.
    \end{equation}
\end{enumerate}

\section{{Design choices and discussion}}
\label{sec6}
\subsection{Parameter tuning}
{The performance of FAMIS is governed primarily by the evolution of the intermediate rare-event targets. The surrogate and tempering schedules should evolve gradually enough to explore the relevant failure regions, while defensive exploration and mixture adaptation preserve coverage of disconnected modes. These parameters should therefore be
tuned jointly.
Practical tuning guidelines are provided below.}

\begin{enumerate}[label=(\alph*), leftmargin=*, align=left]

\item \textit{\textbf{Choice of surrogate threshold and sharpness
($u^{(t)}$ and $\kappa^{(t)}$).}}
Recall the smooth failure surrogate in Eq. \eqref{eq:sigmoid}, where $u^{(t)}$ controls the location of the intermediate failure boundary and $\kappa^{(t)}$ controls its sharpness. We initialize $u^{(t)}>0$ (i.e.,  $u^{(1)}=5$) to enlarge the effective rare-event region and facilitate early exploration, then gradually decrease it to $u^{(t)}=0$ so that the surrogate approaches the true boundary $S(\mathbf{x})=0$. If $u^{(1)}$ is too small, the initial target may be difficult to reach; if it is too large, training effort is spent in regions with little relevance to the final rare event.

{Conversely, the sharpness parameter $\kappa^{(t)}$ is increased during training. Small values provide a smooth surrogate that supports exploration, whereas large values yield a sharper approximation of the failure indicator. Excessively large $\kappa^{(t)}$, however, can make the target nearly discontinuous and destabilize adaptation, particularly for multimodal failure domains. 
In our experiments, $\kappa_{\max}\in [10,25]$ provided a good balance between boundary accuracy and training stability.}

{\item \textit{\textbf{Choice of initial and final temperatures
($\gamma_{\mathrm{init}}$ and $\gamma_{\min}$).}}
The temperature parameter $\gamma^{(t)}$ controls tempering of the intermediate target
    $\pi^{(t)\star}(\mathbf{x})
    \propto
    \left[
        \pi(\mathbf{x})s^{(t)}(\mathbf{x})
    \right]^{1/\gamma^{(t)}}$, as in sequential MC samplers \cite{Moral2006}. A larger initial temperature $\gamma_{\mathrm{init}}$, in Eq. (\ref{gamma_update}). produces a flatter initial target and promotes exploration of separated failure modes, whereas values close to $1$ can lead to premature concentration on a subset of the modes. Excessively large values, however, make the initial and final targets substantially different and can hinder adaptation.
    We set $\gamma_{\mathrm{init}}=20$, which provided good early
exploration of the failure space.}

{The parameter $\gamma_{\min}$ controls the final target concentration. Setting $\gamma_{\min}=1$ recovers the untempered surrogate target, but reaching this value too early can destabilize training when $u^{(t)}$ is already close to zero and $\kappa^{(t)}$ is large. We therefore keep $\gamma^{(t)}>1$ during most of the adaptation and approach $\gamma_{\min}=1$ only toward the end of training.}

{\item \textit{\textbf{Choice of annealing steps and adaptation iterations ($T'$ and $T$).}}
In Eq. \eqref{gamma_update}, $T'$ denotes the number of annealing iterations for $\gamma^{(t)}$, while $T$ is the total number of adaptation iterations. We choose $T'<T$ such that  $\gamma^{(t)}=\gamma_{\min}$ for $t\geq T'$ or before training ends. 
A relatively long annealing period, typically $T'/T\approx0.85$--$0.9$, provided stable adaptation in our experiments, consistent with the observations in \cite{DASGUPTA2024109729}.}  Using \(T\geq10^3\) adaptation iterations generally yield more stable and gradual proposal evolution.

{\item \textit{\textbf{Choice of defensive mixture and KL mixing
weights ($\epsilon^{(t)}$ and $\beta^{(t)}$).}}
The coefficient $\epsilon^{(t)}\in(0,1)$, in Eq. \eqref{eq:defensive_mixture}, denotes the fraction of samples drawn from the defensive density $\pi_0$ in the training mixture $\Psi^{(t)}$. }
{It controls the exploration and exploitation tradeoff: values that are too small may miss uncovered failure modes, whereas large values (e.g., $\epsilon^{(t)}\approx0.5$) can increase gradient noise and slow adaptation. We therefore use a moderate decreasing schedule, from $\epsilon^{(1)}\in [0.3, 0.5]$ to $\epsilon^{(T)}\in [0.05, 0.15]$.}

{The parameter $\beta^{(t)}\in[0,1]$ controls the balance between the reverse-KL refinement term and the forward-KL discovery term in Eq. (\ref{eq:hybrid_kl_fit}).
Small $\beta^{(t)}$ emphasizes local refinement but may miss failure modes, whereas large values promote mode discovery at the cost of a broader, less refined proposal. We therefore use a decreasing schedule, with larger $\beta^{(t)}$ early in training and smaller values later, e.g.,
$\beta^{(1)}=0.5$ and 
$\beta^{(T)}=0.1$. Together, the decreasing schedules prioritize broad exploration and mode discovery early in training, followed by proposal refinement as adaptation progresses.}

\item \textit{\textbf{Choice of the number of components and batch size (\(N\) and $K$).}}
The batch size $K$ controls the number of samples drawn from the defensive mixture at each iteration. Larger $K$ reduces the variability of the stochastic updates but increases the number of LSF evaluations. 
In our experiments, \(K=100\) provided satisfactory performance.

The number of components controls mixture flexibility and computational cost, but not the number of LSF calls. Small mixtures can be sufficient for relatively simple failure geometries, while additional components may improve coverage of separated regions. It is straightforward to choose to choose the number of proposal components $N$ when the number of failure modes is known.  In problems with nearly unimodal or moderately nonlinear failure region, a small number of components, i.e., $N=1$, is usually sufficient. For highly disconnected failure regions, $N>1$ will ensure the mixture accurately covers all important modes. 
However, for a limited batch size \(K\), choosing a smaller \(N\) is preferable to maintain the quality and stability of the parameter estimates.
\end{enumerate}

\subsection{Discussion}


Because all proposal components share the same invertible transformation
$f_\phi$,  $\operatorname{JSD}(q_n^{(t)} \| q_\alpha^{(t)})
=
\operatorname{JSD} (q_{0,n}^{(t)} \| q_{0,\alpha}^{(t)})$, where $q_{0,\alpha}^{(t)}
=
\sum_{n=1}^{N}\alpha_n^{(t)}q_{0,n}^{(t)}$. 
Hence, for fixed base parameters and mixture weights,
$\nabla_\phi \mathcal{R}_{\mathrm{JSD}}^{(t)}=0$.
As a result, the JSD term directly promotes separation among the
base components, while the hybrid fit objective in Eq. (\ref{eq:full_objective_with_jsd}) trains the shared flow.
The training procedure also avoids requiring derivatives of the LSF.
At each iteration, FAMIS first samples from the current defensive mixture and evaluates the LSF at those points. The proposal is then updated using these evaluations, without requiring gradients of the LSF or differentiating through the sampling step. This makes FAMIS applicable to black-box reliability models for which input gradients are unavailable.

Another key advantage of FAMIS is that all mixture components share a common normalizing flow $f_{{\phi}}$, parameterized by ${\phi}$, rather than using a separate flow network for each proposal. The shared flow captures the global nonlinear geometry of the rare-event target, while the distinct Gaussian bases $q_{0,n}$, with $\boldsymbol{\theta}_n=(\boldsymbol{\mu}_n,\boldsymbol{\Sigma}_n)$, allow the induced proposals $q_n(\mathbf{x};\boldsymbol{\theta}_n,\boldsymbol{{\phi}})$ to represent different local failure regions. This structure reuses both the flow transformation and a common batch of model evaluations, so increasing $N$ improves mixture flexibility without increasing $N_{\mathrm{call}}$, although density-evaluation and optimization costs grow with $N$. Overall, the shared-flow design provides a computationally efficient way to build a flexible multimodal proposal, particularly when LSF evaluations are expensive.

\section{Numerical examples}
\label{sec7}

To assess the accuracy and efficiency of FAMIS in estimating very small failure probabilities, five examples are considered. Example~1 uses toy multimodal targets to test mode recovery under different weights and tail behaviors. Example~2 considers two-dimensional truncated-tail events to illustrate adaptation to disconnected, curved, and nonconvex failure regions. Example~3 studies a non-Gaussian quadratic reliability problem with correlated Gumbel marginals. 
Example~4 examines scalability on a high-dimensional benchmark with a highly nonlinear LSF. Example~5 considers a high-dimensional hyperspherical failure event under Neal's funnel distribution to test performance under strong non-Gaussian dependence and geometric distortion.

In the benchmark examples, results are compared with the state-of-the-art REIN \cite{DASGUPTA2024109729}, approximate sampling target with post-processing adjustment (ASTPA) \cite{ESHRA2025111200} method, and component-wise Metropolis–Hastings based subset simulation (CWMH-SS) \cite{AU2001263}. For CWMH-SS, we use $K=5\times10^4$ samples per subset level to obtain a stable high-dimensional baseline.
  For all test cases, reference values were obtained using direct MC simulation with \(10^{10}\) samples. For Examples~2 and 3, performance is measured by the relative root mean-square error (RRMSE) of \(\widehat{P}_f\), averaged over 100 independent runs. For Examples~4 and 5, we report the relative mean absolute logarithmic error (MALE) to account for underestimation and overestimation penalties in high dimensions.

\subsection{{NF architecture and training.}}
{For all following examples, we use a normalizing flow model (RealNVP) with $L=6$ affine coupling layers. Each coupling layer employs a two hidden layers with 128 neurons per layer and ReLU activations.
All flow and component parameters are optimized using Adam with learning rate $\eta=0.002$ over $T=700$ iterations with $K_{\text{est}}=2\times10^{3}$. The repulsion coefficient is fixed at
$\lambda_{\mathrm{JSD}} = \text{2}$ in all experiments.}

\subsection{Evaluation criteria}
\label{evaluation_criteria}

The performance of the estimators is assessed over \(R=100\) independent runs. Let
\(\widehat{P}_f^{(i)}\) denote the estimate from the \(i\)-th run and let \(P_f\) be the reference failure probability. We report the empirical mean estimate
\begin{equation}
\mathbb{E}[\widehat{P}_f]
\approx
\frac{1}{R}\sum_{i=1}^{R}\widehat{P}_f^{(i)} .
\end{equation}
The estimator variability is measured by the coefficient of variation,
\begin{equation}
\delta_{\widehat{P}_f}(\%)
=
100\,
\frac{
\sqrt{\widehat{\mathrm{Var}}(\widehat{P}_f)}
}{
\mathbb{E}[\widehat{P}_f]
},
\end{equation}
where $\widehat{\mathrm{Var}}(\widehat{P}_f)
=
\frac{1}{R-1}
\sum_{i=1}^{R}
\left(
\widehat{P}_f^{(i)}
-
\mathbb{E}[\widehat{P}_f]
\right)^2$ is the sample variance across the \(R\) independent runs. 
Accuracy is quantified using the relative root mean-square error,
{\small
\begin{equation}
\mathrm{RRMSE}
=
\sqrt{
\frac{1}{R}\sum_{i=1}^{R}
\left(
\frac{\widehat{P}_f^{(i)}-P_f}{P_f}
\right)^2
}
\approx
\sqrt{
\varepsilon_{\mathrm{rel}}^2
+
\left(
\frac{\mathbb{E}[\widehat{P}_f]}{P_f}
\right)^2
\delta_{\widehat{P}_f}^{\,2}
},
\end{equation}
}
where 
$\varepsilon_{\mathrm{rel}}
=
\frac{\mathbb{E}[\widehat{P}_f]-P_f}{P_f}$ is the relative bias. For examples involving extremely small probabilities, we also report the mean absolute logarithmic error (MALE),
\begin{equation}
\mathrm{MALE}
=
\frac{1}{R}\sum_{i=1}^{R}
\left|
\log\left(
\frac{\widehat{P}_f^{(i)}}{P_f}
\right)
\right|,
\end{equation}
which penalizes both underestimation and overestimation on a logarithmic scale. The mean of the evaluation metrics is reported in the relevant tables. All methods were carefully tuned for each example to ensure a fair comparison and their best observed performance.

\begin{figure*}[!t]
    \centering
    \begin{subfigure}[t]{0.32\textwidth}
        \centering
        \includegraphics[width=\linewidth]{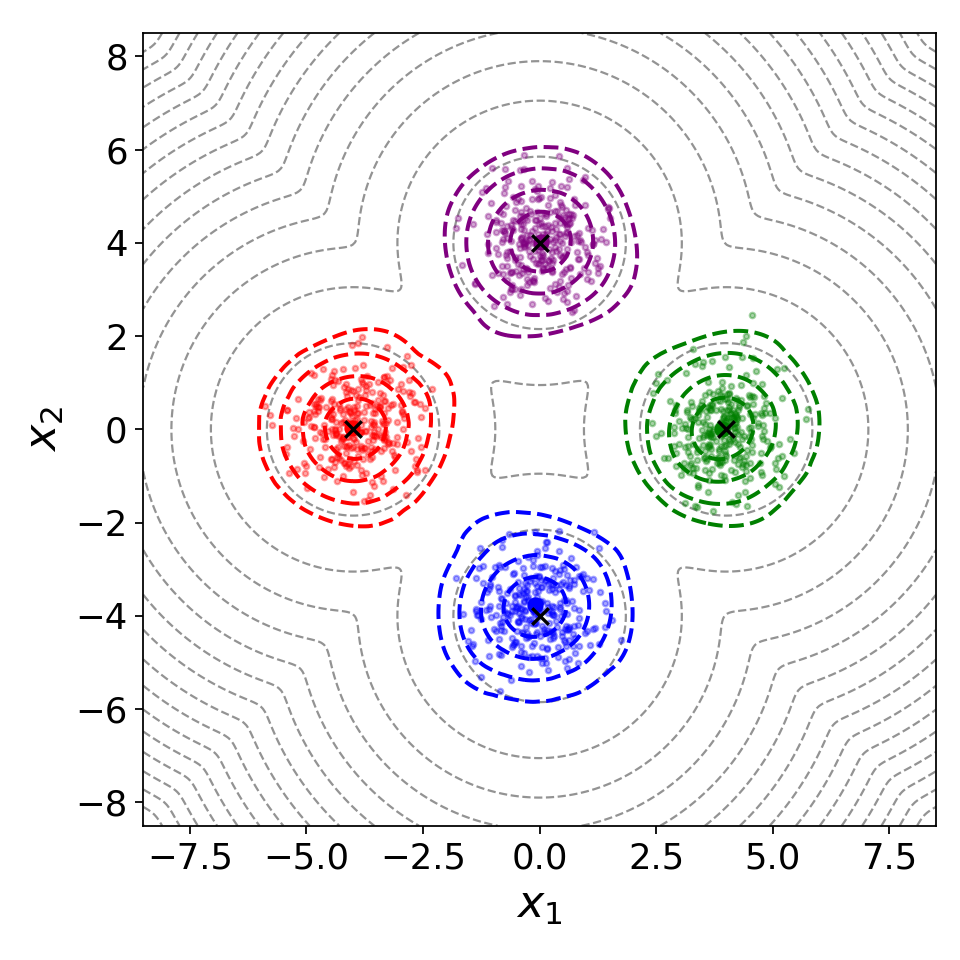}
        \caption{}
        \label{fig:one}
    \end{subfigure}\hfill
    \begin{subfigure}[t]{0.32\textwidth}
        \centering
        \includegraphics[width=\linewidth]{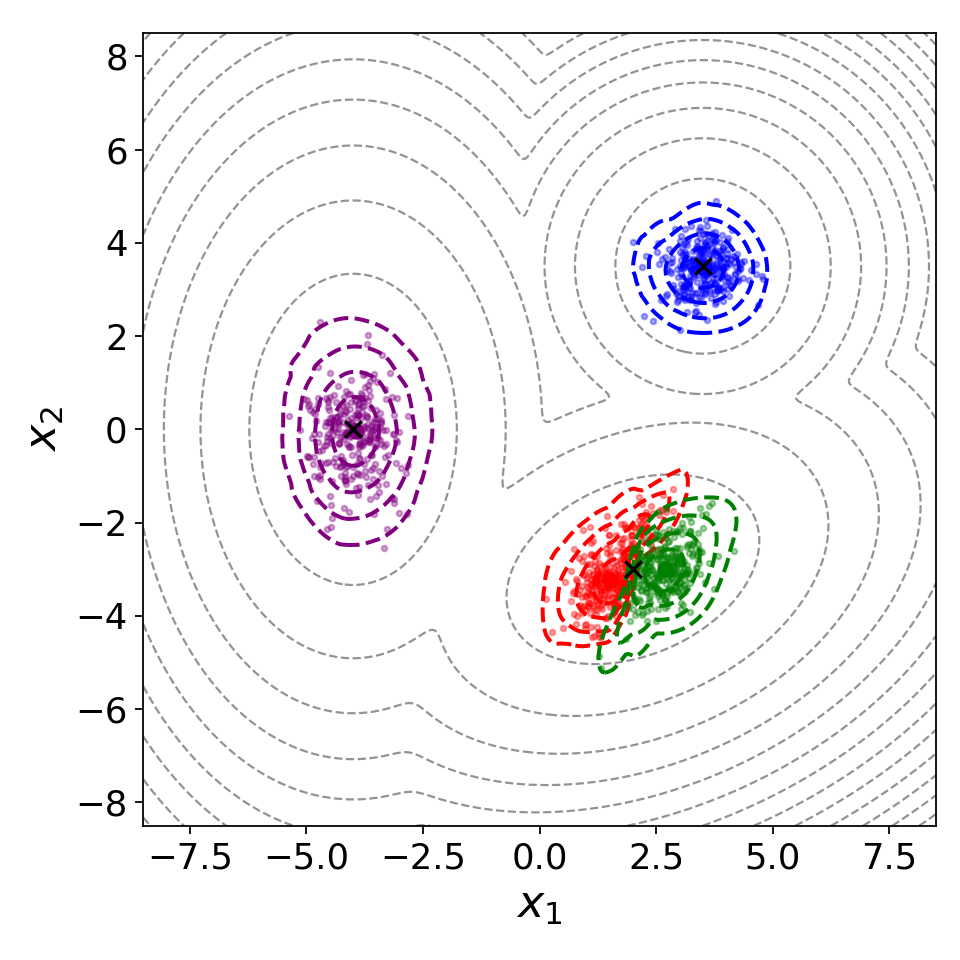}
        \caption{}
        \label{fig:two}
    \end{subfigure}\hfill
    \begin{subfigure}[t]{0.32\textwidth}
        \centering
        \includegraphics[width=\linewidth]{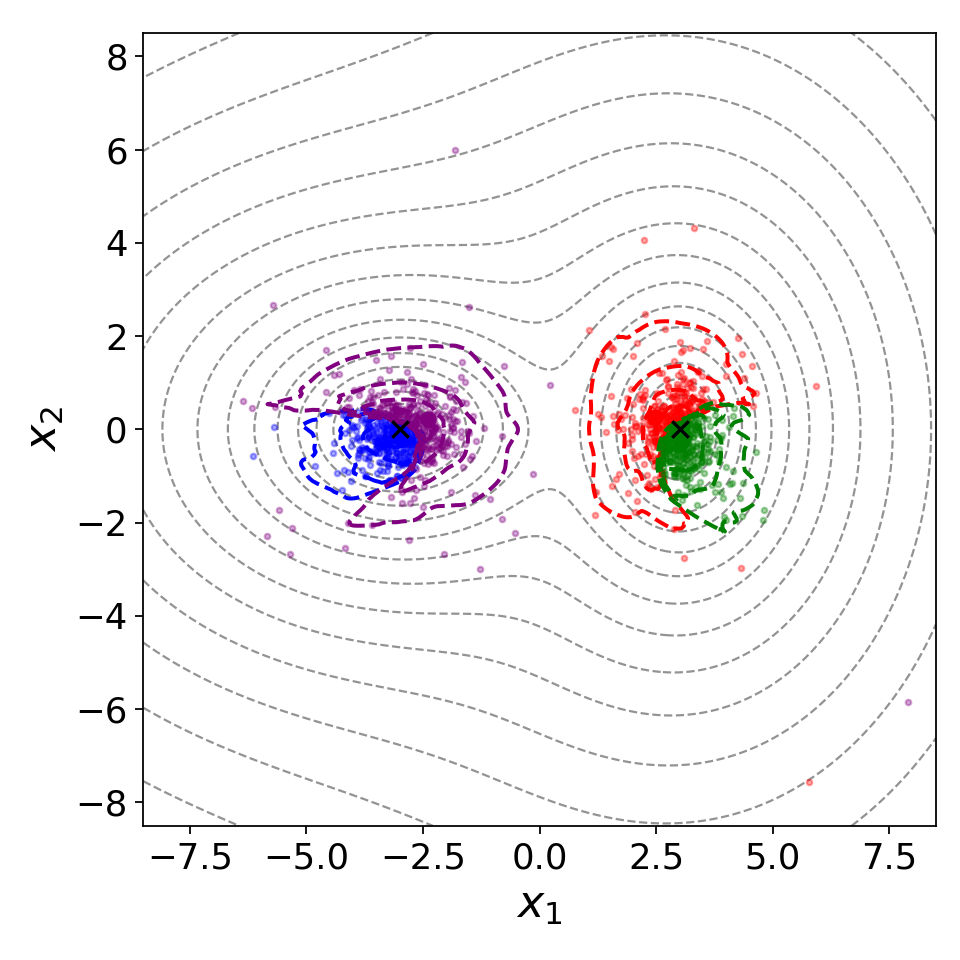}
        \caption{}
        \label{fig:three}
    \end{subfigure}

  \caption{\textbf{Example~1.} Toy multimodal targets and learned $N=4$ NF proposals ($d_x=2$). 
Contours show the target log-density $\log {\pi}(\mbf{x})$ and markers indicate the target mode locations. 
(a) Symmetric four-mode Gaussian-mixture target with the learned NF proposals capturing all modes.
(b) Imbalanced three-mode Gaussian-mixture target; the learned proposal allocates distinct components across the unequal mode structure without component collapse.
(c) Two-mode Student's $t$ mixture target with heavy tails; the learned proposal concentrates around both modes while maintaining adequate tail coverage.}
\label{fig:toy_targets}
\end{figure*}

\subsection{Example 1: Toy multimodal targets}
To qualitatively evaluate how well the proposed flow mixture proposals capture multimodality, we consider
three two-dimensional synthetic unnormalized target densities $\pi(\mathbf{x})$ on $\mathbb{R}^2$. 
We set the batch size $K=100$ and the scaling constant
$c=2$, and define three synthetic multimodal targets:

\paragraph{Equally weighted Gaussian mixture}  The target $\pi$ is a symmetric mixture density of four
equally weighted Gaussian distributions such that
\begin{equation}
{\pi}(\mathbf{x})
=
\frac{1}{4}
\sum_{k=1}^{4}
\mathcal{N}\!\left(\mathbf{x};\,\boldsymbol{\mu}_k,\,0.7^2 I_2\right),
\label{eq:toy_gauss4_equal_compact}
\end{equation}
where $\boldsymbol{\mu}_k
\in
\left\{
(-4,0)^\top,\,
(4,0)^\top,\,
(0,4)^\top,\,
(0,-4)^\top
\right\}$ and $I_2$ is the $2\times 2$ identity matrix. This example tests whether the flow mixture $q_{\alpha}$ can allocate distinct components to several
disjoint modes with equal importance.

\paragraph{ Imbalanced anisotropic Gaussian mixture} The target $\pi$ is a mixture density of three
two-dimensional Gaussian distributions with unequal weights and multiplied by c such that
\begin{equation}
\begin{aligned}
    {\pi}(\mathbf{x})=
c\Big[
0.55\,\mathcal{N}\!\big(\mathbf{x};\,\boldsymbol{\mu}_1,\,\boldsymbol{\Sigma}_1\big)
&+0.30\,\mathcal{N}\!\big(\mathbf{x};\;\,\boldsymbol{\mu}_2,\,\boldsymbol{\Sigma}_2\big)\\
+0.15\,\mathcal{N}\!\big(\mathbf{x};\;\,\boldsymbol{\mu}_3,\,\boldsymbol{\Sigma}_3\big)
\Big],
\end{aligned}
\end{equation}
with $\boldsymbol{\mu}_1=(-4,0)^\top,
\boldsymbol{\mu}_2=(3.5,3.5)^\top,
\boldsymbol{\mu}_3=(2,-3)^\top$ and $\boldsymbol{\Sigma}_1=\operatorname{diag}(0.6^2,0.9^2)$, 
$\boldsymbol{\Sigma}_2=0.5^2 I_2$,
 $\boldsymbol{\Sigma}_3=
\begin{pmatrix}
0.8^2 & 0.14\\
0.14 & 0.6^2
\end{pmatrix}$. This example tests whether the learned $q_{\alpha}$ can adapt to unequal modal contributions
and anisotropic local covariance structures.

\paragraph{Equally weighted Student's $t$ mixture}
Let $t_\nu(\mathbf{x};\boldsymbol{\mu},\boldsymbol{\Sigma})$ denote the multivariate Student's $t$ density with $\nu$ degrees of freedom,
location $\boldsymbol{\mu}$, and scale matrix $\boldsymbol{\Sigma}$. With $\nu=2$,
\begin{equation}
{\pi}(\mathbf{x})
=
2\left[
\frac{1}{2}t_{\nu}\!\left(\mathbf{x};\boldsymbol{\mu}_1,\boldsymbol{\Sigma}_1\right)
+
\frac{1}{2}t_{\nu}\!\left(\mathbf{x};\boldsymbol{\mu}_2,\boldsymbol{\Sigma}_2\right)
\right],
\label{eq:toy_student2}
\end{equation}
where $\boldsymbol{\mu}_1=(-3,0)^\top$, $\boldsymbol{\mu}_2=(3,0)^\top$, and $\boldsymbol{\Sigma}_1=\operatorname{diag}(0.8^2,0.6^2)$, $\boldsymbol{\Sigma}_2=\operatorname{diag}(0.6^2,0.8^2)$. This example tests robustness to heavier tails while retaining a multimodal structure.

As shown in Fig.~\ref{fig:toy_targets}(a), the learned mixture of flow proposals recovers the symmetric multimodal
structure of the equally-weighted Gaussian mixture, with distinct components covering the separated modes.
Fig.~\ref{fig:toy_targets}(b) demonstrates that, under an imbalanced Gaussian mixture, the proposal adapts its
mass allocation to match the unequal contributions of the modes while the JSD repulsion term mitigates component
collapse. Finally, Fig.~\ref{fig:toy_targets}(c) considers a heavy-tailed two-component Student's $t$ mixture,
where the learned proposal captures both modes and preserves broader support consistent with the target tails.

\subsection{Example 2: Qualitative truncated-tail examples (2D)}

\begin{figure*}[t]
\centering

\begin{subfigure}[t]{0.19\textwidth}
    \centering
    \includegraphics[width=\linewidth,height=3.0cm]{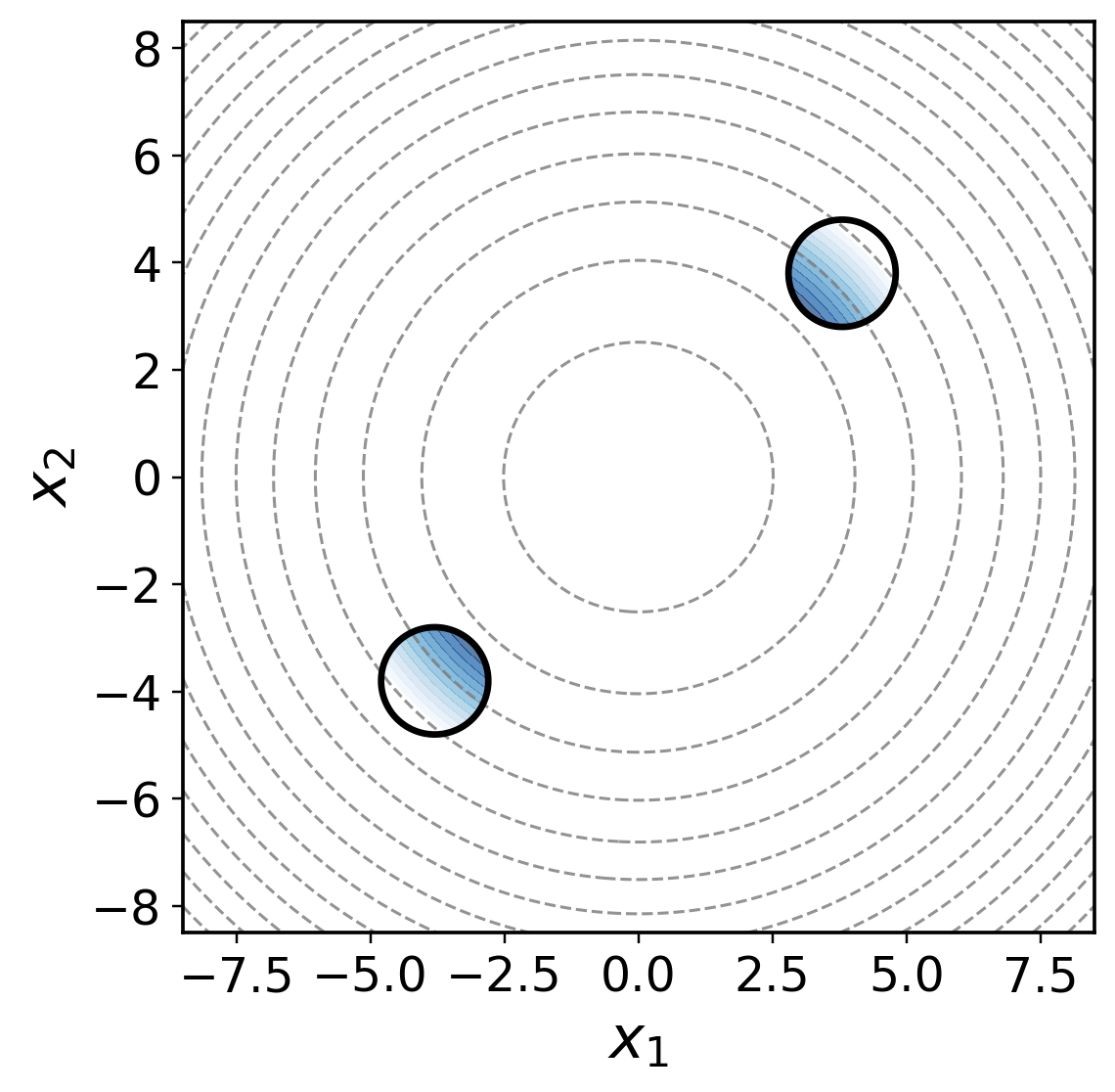}

    \vspace{0.25cm}

    \includegraphics[width=\linewidth,height=3.0cm]{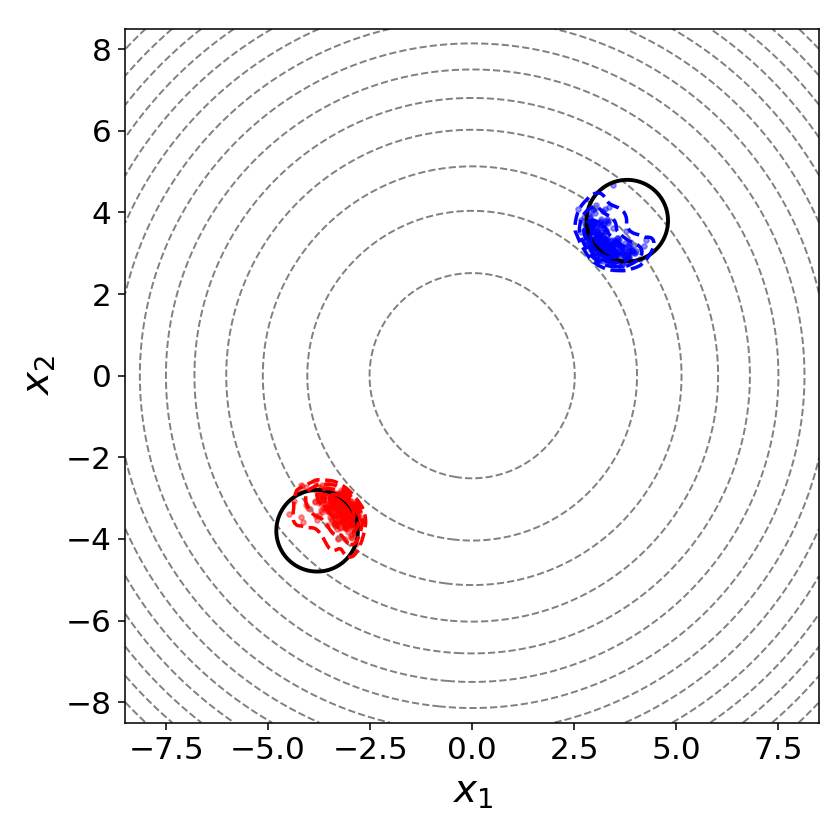}
    \caption{}
    \label{fig:a}
\end{subfigure}\hfill
\begin{subfigure}[t]{0.19\textwidth}
    \centering
    \includegraphics[width=\linewidth,height=3.0cm]{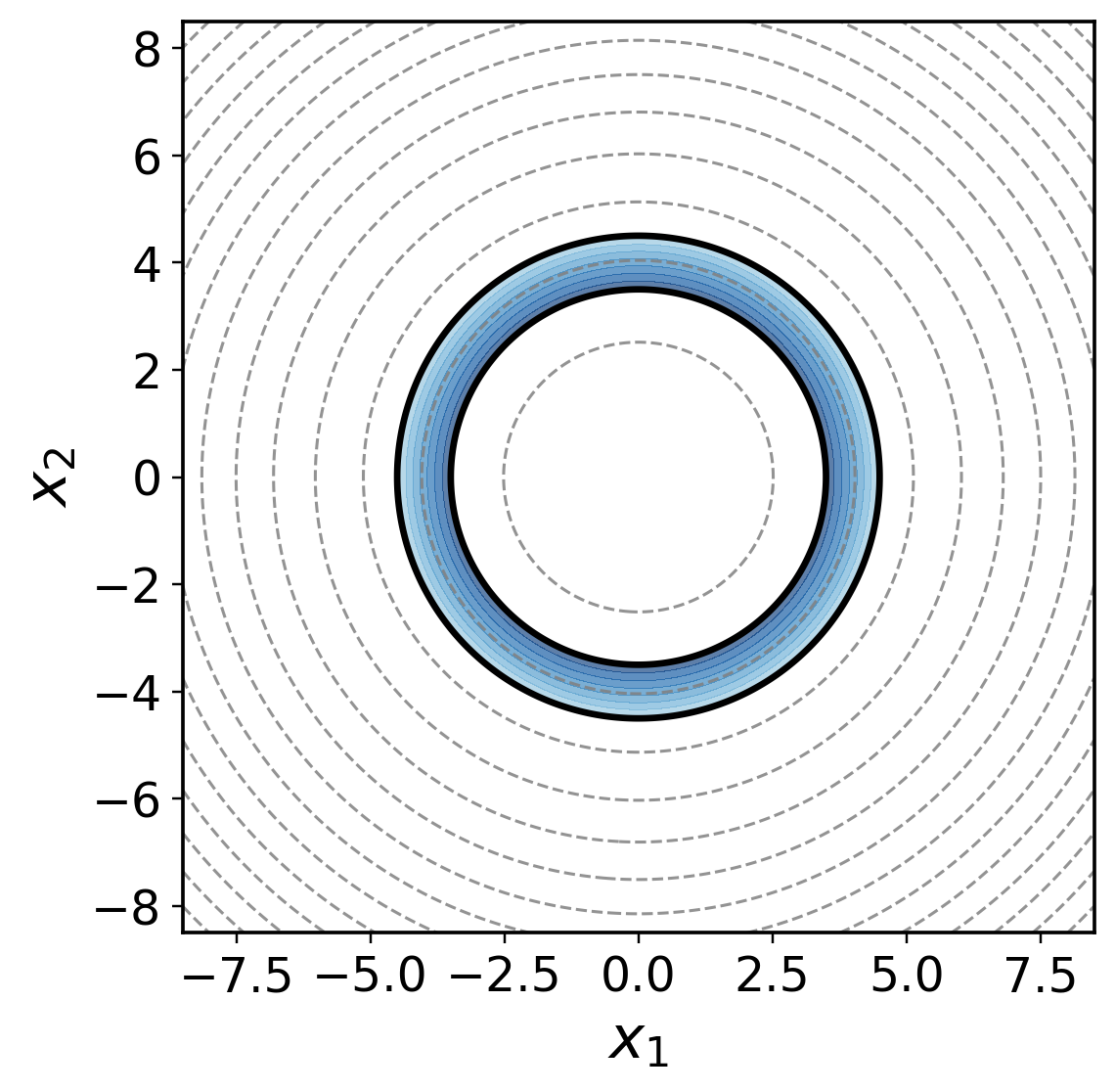}

    \vspace{0.25cm}

    \includegraphics[width=\linewidth,height=3.0cm]{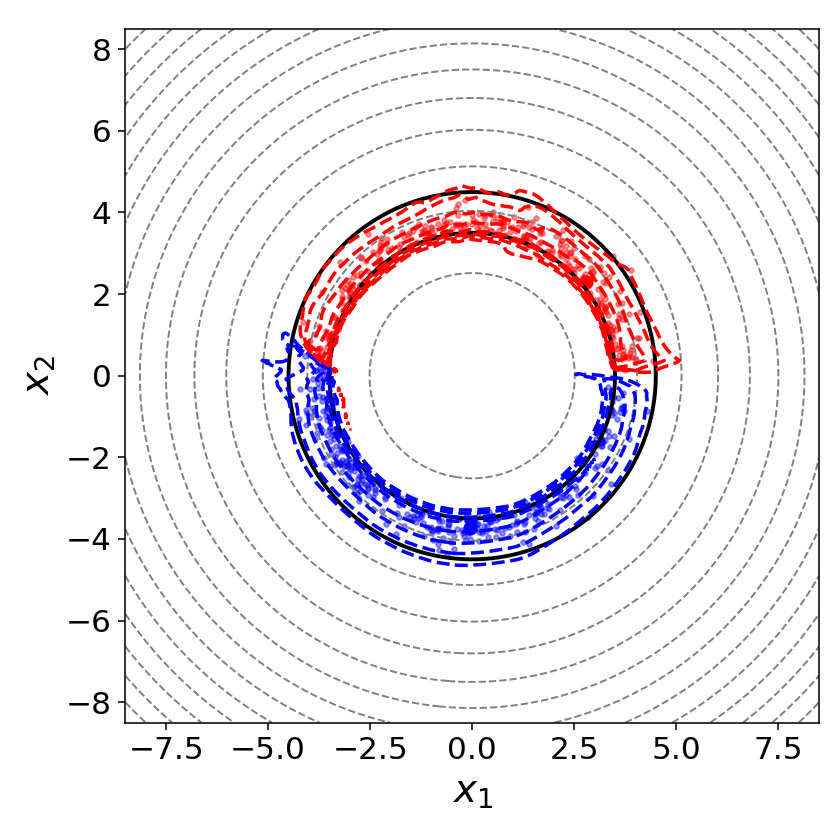}
    \caption{}
    \label{fig:b}
\end{subfigure}\hfill
\begin{subfigure}[t]{0.19\textwidth}
    \centering
    \includegraphics[width=\linewidth,height=3.0cm]{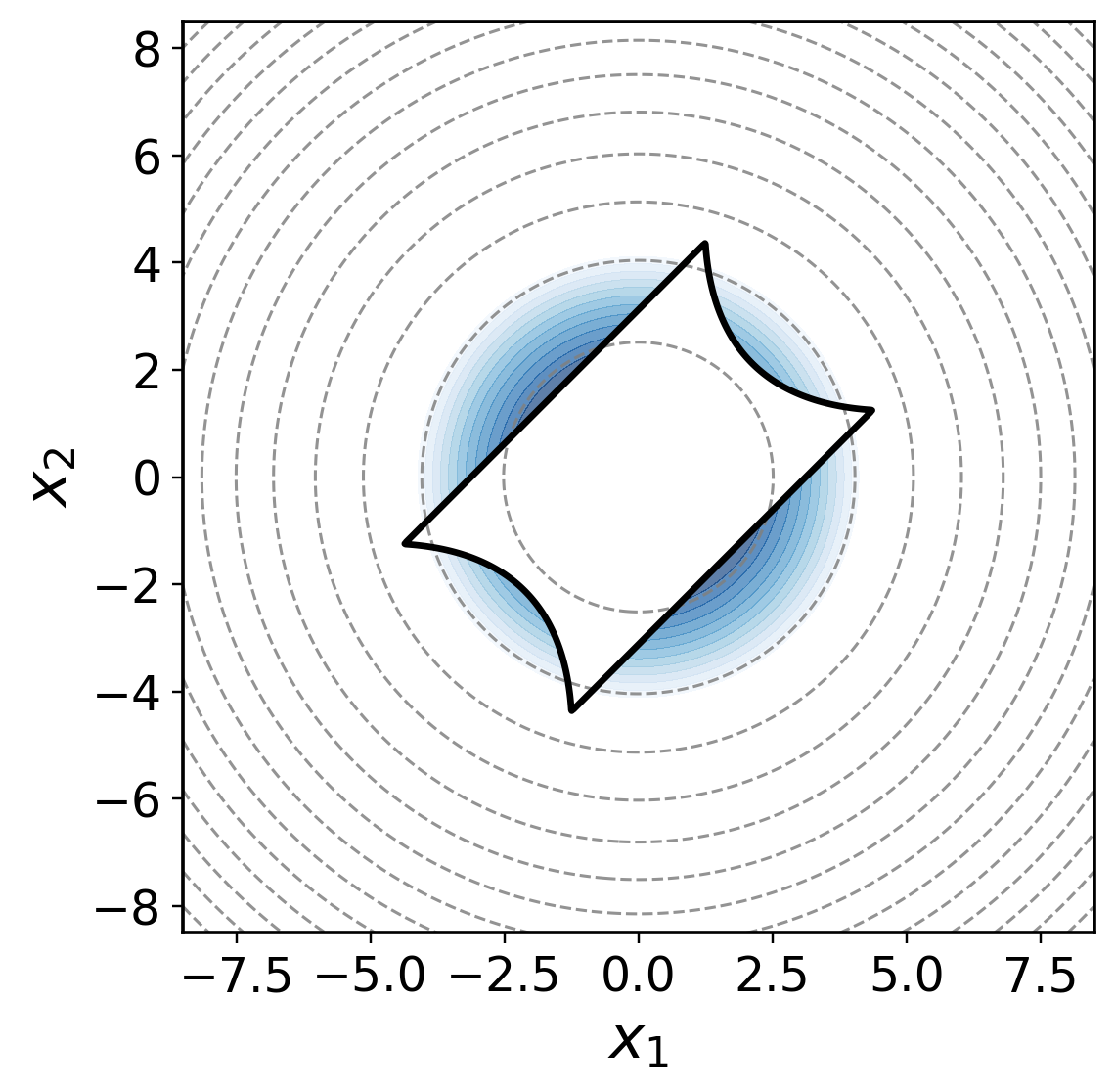}

    \vspace{0.25cm}

    \includegraphics[width=\linewidth,height=3.0cm]{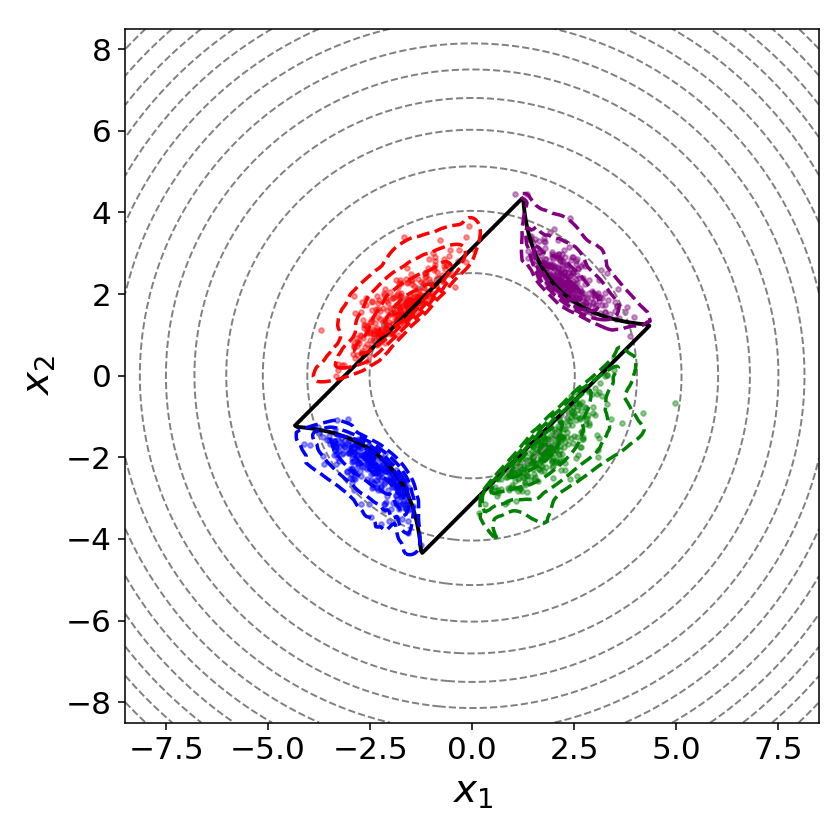}
    \caption{}
    \label{fig:c}
\end{subfigure}\hfill
\begin{subfigure}[t]{0.19\textwidth}
    \centering
    \includegraphics[width=\linewidth,height=3.0cm]{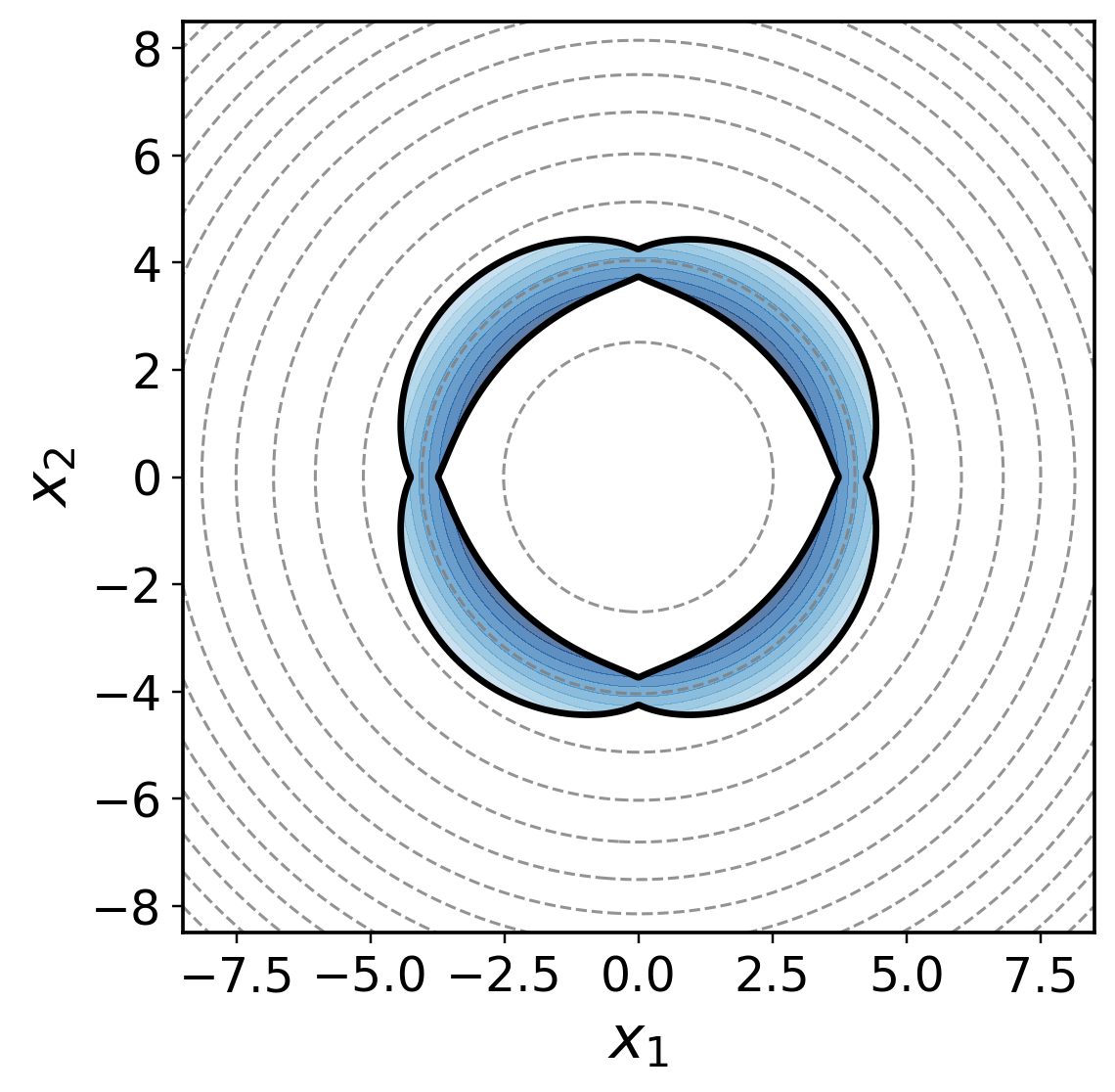}

    \vspace{0.25cm}

    \includegraphics[width=\linewidth,height=3.0cm]{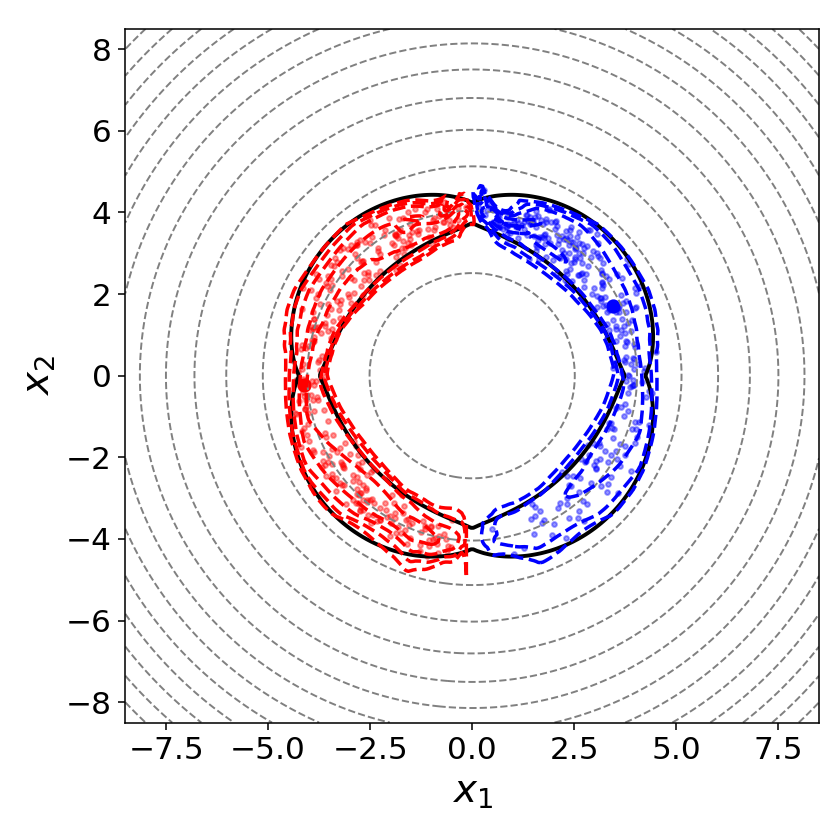}
    \caption{}
    \label{fig:d}
\end{subfigure}\hfill
\begin{subfigure}[t]{0.19\textwidth}
    \centering
    \includegraphics[width=\linewidth,height=3.0cm]{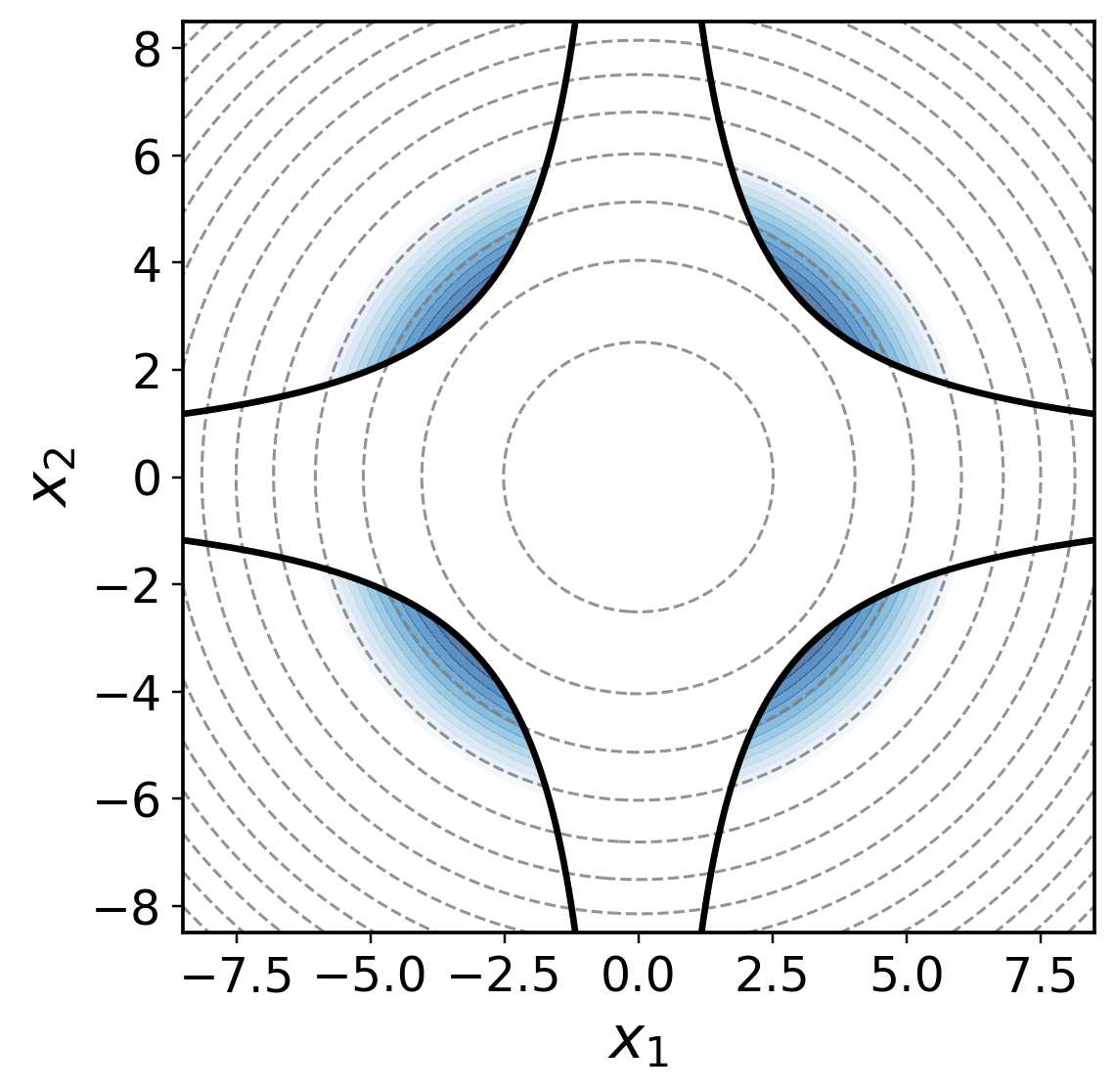}

    \vspace{0.25cm}

    \includegraphics[width=\linewidth,height=3.0cm]{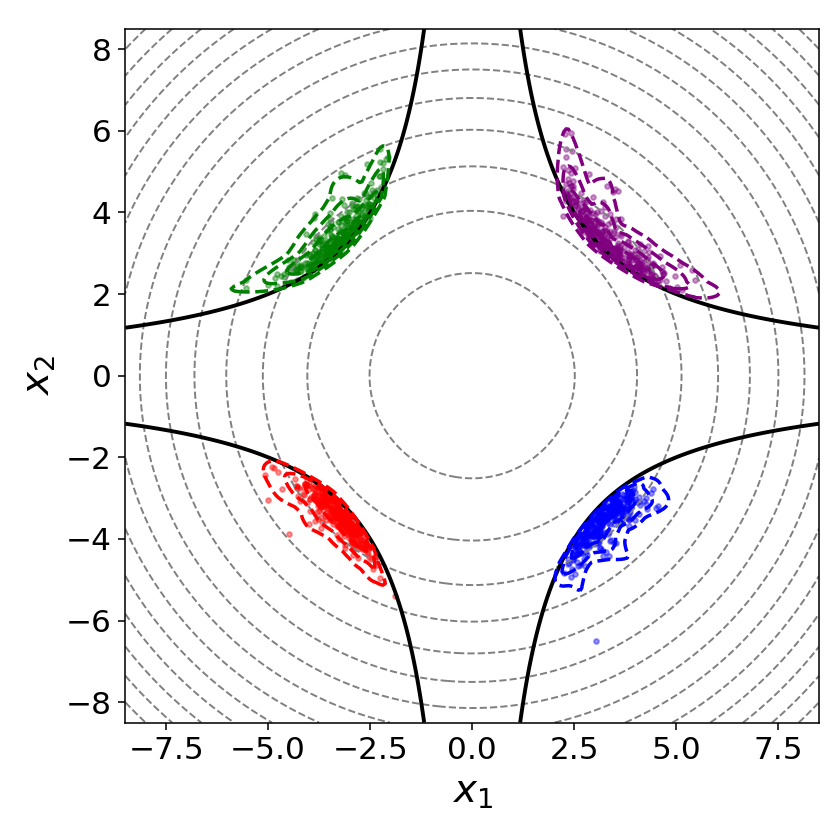}
    \caption{}
    \label{fig:e}
\end{subfigure}

\caption{\textbf{Example~2.} Qualitative truncated-tail targets with a nominal density $\widetilde{\pi}(\mbf{x})=\mathcal{N}(\mathbf{0},I_2)$.
(a)--(e) Top: truncated optimal proposal $q^\star(\mbf{x})\propto \widetilde{\pi}(\mbf{x})\mathbb{I}\{\mbf{x}\in\mathcal{F}\}$.
Bottom: learned mixture proposal $q_\alpha$ produced by FAMIS.}
\label{fig:qualitative_2d}
\end{figure*}
We consider several two-dimensional rare-event examples adapted from the qualitative
benchmarks in \cite{gao2024nofis}. In all cases, the target density is a standard bivariate Gaussian density
\(\widetilde{\pi}(\mathbf{x})=\mathcal{N}(\mathbf{0},\mathbf{I}_2)\), with
\(\mathbf{x}=(x_1,x_2)^\top\), and the failure domain is defined as
    $\mathcal{F}=\{\mathbf{x}\in\mathbb{R}^2:S(\mathbf{x})\leq 0\}$.
The five truncation sets are chosen to cover multimodal and nonconvex geometries:   
\begin{table*}[!b]
\centering
\caption{Example 2:  Empirical RRMSE for the LSFs in Eqs. (\ref{eq:g_case_b})-(\ref{eq:g_case_g}) across
different training batch size $K $ and number of proposals $N$.}
\label{tab:lsf_K_N_results}
\renewcommand{\arraystretch}{1.55}
\setlength{\tabcolsep}{8pt}
\begin{tabular}{c cc cc cc cc}
\toprule
\toprule
\multirow{2}{*}{LSF}
& \multicolumn{2}{c}{$K=50$}
& \multicolumn{2}{c}{$K=100$}
& \multicolumn{2}{c}{$K=500$}
& \multicolumn{2}{c}{$K=1000$} \\
\cmidrule(lr){2-3}\cmidrule(lr){4-5}\cmidrule(lr){6-7}\cmidrule(lr){8-9}
& $N=2$ & $N=4$
& $N=2$ & $N=4$
& $N=2$ & $N=4$
& $N=2$ & $N=4$ \\
\midrule

$S_a$
& $0.017$ & $0.015$
& $0.014$  & $0.010$
& $0.006$ & $0.004$
& $0.002$ & $0.002$ \\

\midrule

$S_b$
& $0.044$ & $0.068$
& $0.039$ & $0.032$
& $0.019$ & $0.018$
& $0.003$ & $0.004$ \\

\midrule

$S_c$
& $0.510$ & $0.033$
& $0.491$ & $0.021$
& $0.420$ & $0.015$
& $0.381$ & $0.014$ \\

\midrule

$S_d$
& $0.142$ & $0.124$
& $0.090$ & $0.103$
& $0.072$ & $0.065$
& $0.035$ & $0.018$ \\

\midrule

$S_e$
& $0.083$ & $0.182$
& $0.062$ & $0.050$
& $0.030$ & $0.027$
& $0.048$ & $0.026$ \\

\bottomrule
\bottomrule
\end{tabular}
\end{table*}
\begin{itemize}

\item \textit{Two-disk tail event (a).}
{\small
\begin{equation}
S_a(\mathbf{x})
=
\min\!\Big\{
(x_1+3.8)^2+(x_2+3.8)^2,\;
(x_1-3.8)^2+(x_2-3.8)^2
\Big\}-1 .
\label{eq:g_case_b}
\end{equation}
}
Hence, \(\mathcal{F}\) is the union of two unit-radius disks centered at
\((-3.8,-3.8)\) and \((3.8,3.8)\), which lie in the tails of $\pi(\mathbf{x})$. The reference value of failure probability is $P_f= 4.7934\times 10^{-6}$.

\item \textit{Annulus (ring) event (b).}
\begin{equation}
S_b(\mathbf{x})
=
\left|\sqrt{x_1^2+x_2^2}-4\right|-0.5 .
\label{eq:g_case_c}
\end{equation}
This defines a ring-shaped failure region centered at the origin, with radius approximately
\(4\) and thickness \(1\), i.e.,
   $ \mathcal{F}
    =
    \left\{\mathbf{x}: 3.5 \leq \sqrt{x_1^2+x_2^2} \leq 4.5\right\}.$ The corresponding failure probability is $P_f= 2.1474\times 10^{-3}$.
\item \textit{Product tail event (c).}
\begin{equation}
S_c(\mathbf{x})
=
10-\left|x_1x_2\right|.
\label{eq:g_case_f}
\end{equation}
Therefore, $\mathcal{F}
    =
    \left\{\mathbf{x}: |x_1x_2|\geq 10\right\}.$
This produces four disconnected tail regions, one in each quadrant. The reference failure probability is $P_f= 1.0832\times 10^{-5}$.

\item \textit{Four-petal event (d).}

\begin{equation}
S_d(\mathbf{x})
=
\min\!\Big\{
(r^2-16)^2+4x_1x_2,\;
(r^2-16)^2-4x_1x_2
\Big\}-4,
\label{eq:g_case_e}
\end{equation}
where \(r^2=x_1^2+x_2^2\). The corresponding failure domain consists of four curved, symmetric tail regions arranged around
the circle \(r\approx 4\). This example is useful for assessing whether the adaptive mixture can
cover nonconvex failure region. The reference failure probability is $P_f= 2.9564\times 10^{-3}$.

\item \textit{Four-branch parabolic event (e).}
\begin{equation}
\begin{aligned}
S_e(\mathbf{x})
=
\min\Bigg\{&
a+\frac{(x_1-x_2)^2}{10}-\frac{x_1+x_2}{\sqrt{2}},\\
&
a+\frac{(x_1-x_2)^2}{10}+\frac{x_1+x_2}{\sqrt{2}},\\
&
(x_1-x_2)+\frac{b}{\sqrt{2}}+1,\\
&
(x_2-x_1)+\frac{b}{\sqrt{2}}+1
\Bigg\},
\end{aligned}
\label{eq:g_case_g}
\end{equation}
for \(a=3\) and \(b=3\).
This case defines a nonconvex four-branch failure region. It is more challenging than the disk or
annulus examples because the failure set contains several elongated branches with different local
geometries. The reference failure probability is $P_f= 2.9063\times 10^{-2}$.
\end{itemize}

Fig.~\ref{fig:qualitative_2d} summarizes the behavior of our proposed algorithm on these examples. The central circular contours represent the target distribution. We use $N=2$ proposal components for examples (a), (b), and (d), and $N=4$ for examples (c) and (e). In panels~(a)--(e), the {top} row of Fig.~\ref{fig:qualitative_2d} visualizes the optimal sampling distribution $q^\star$, while the {bottom} row shows the learned mixture proposal $q_\alpha$ returned by our algorithm.

 Across all cases, the bottom row shows that the learned mixture proposal $q_\alpha(x)$ concentrates probability mass on the same tail geometry as $q^\star$. In particular, in the multimodal case~(a), the mixture allocates mass to both disconnected disks without collapsing
to a single mode, while in~(b)--(e) it captures nonconvex tubular and annular geometries.

\begin{table*}[!b]
\centering
\caption{\textbf{Example 3:} Performance results of FAMIS for highly correlated Gumbel input density with $N=1$.}
\label{tab:correlated_gumbel_results}
\renewcommand{\arraystretch}{1.55}
\resizebox{\linewidth}{!}{%
\begin{tabular}{lccccccc}
\toprule
\toprule
& & \multicolumn{3}{c}{$K=100$} & \multicolumn{3}{c}{$K=500$} \\
\cmidrule(lr){3-5}\cmidrule(lr){6-8}
& $P_f$
& $\mathbb{E}[\widehat{P}_f]$
& $\delta_{\widehat{P}_f}(\%)$
& RRMSE
& $\mathbb{E}[\widehat{P}_f]$
& $\delta_{\widehat{P}_f}(\%)$
& RRMSE \\
\midrule

$d_x=2$ 
& & & & & & & \\
$\lambda=70$
& $2.51\times 10^{-7}$
& $2.48\times 10^{-7}$
& $7.62$
& $0.075$
& $2.51\times 10^{-7}$
& $1.22$
& $0.012$ \\
$m=2$
& & & & & & & \\

\midrule

$d_x=3$
& & & & & & & \\
$\lambda=5$
& $4.23\times 10^{-7}$
& $4.21\times 10^{-7}$
& $6.21$
& $0.062$
& $4.24\times 10^{-7}$
& $4.12$
& $0.041$ \\
$m=3$
& & & & & & & \\

\midrule

$d_x=40$
& & & & & & & \\
$\lambda=-200$
& $4.60\times 10^{-6}$
& $4.49\times 10^{-6}$
& $12.30$
& $0.120$
& $4.54\times 10^{-6}$
& $7.49$
& $0.074$ \\
$m=20$
& & & & & & & \\

\bottomrule
\bottomrule
\end{tabular}
}
\end{table*}
\begin{figure*}[!b]
    \centering
    \begin{subfigure}[t]{0.3\textwidth}
        \centering
        \includegraphics[width=\linewidth]{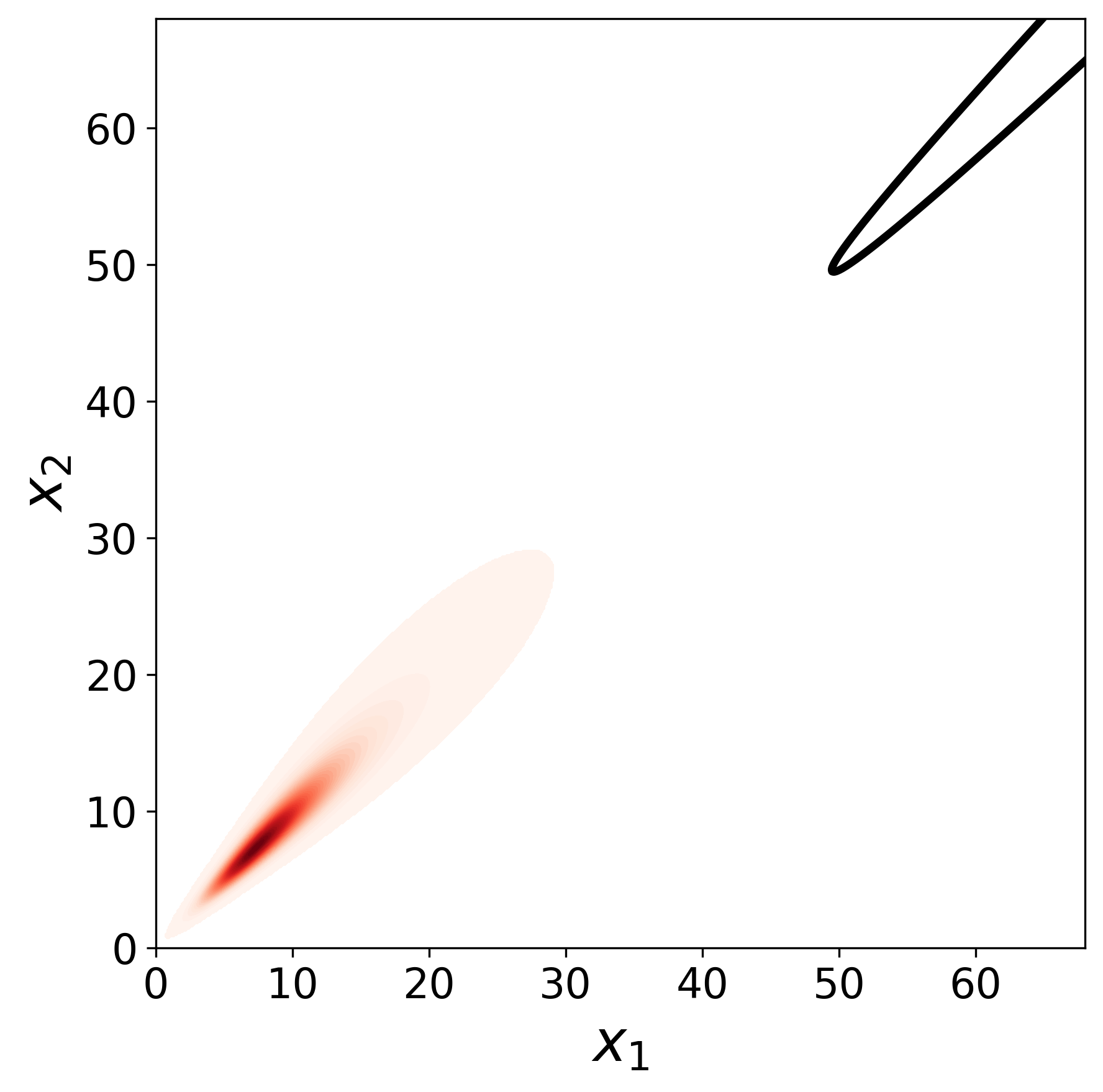}
        \caption{}
        \label{fig:one}
    \end{subfigure}\hfill
    \begin{subfigure}[t]{0.3\textwidth}
        \centering
        \includegraphics[width=\linewidth]{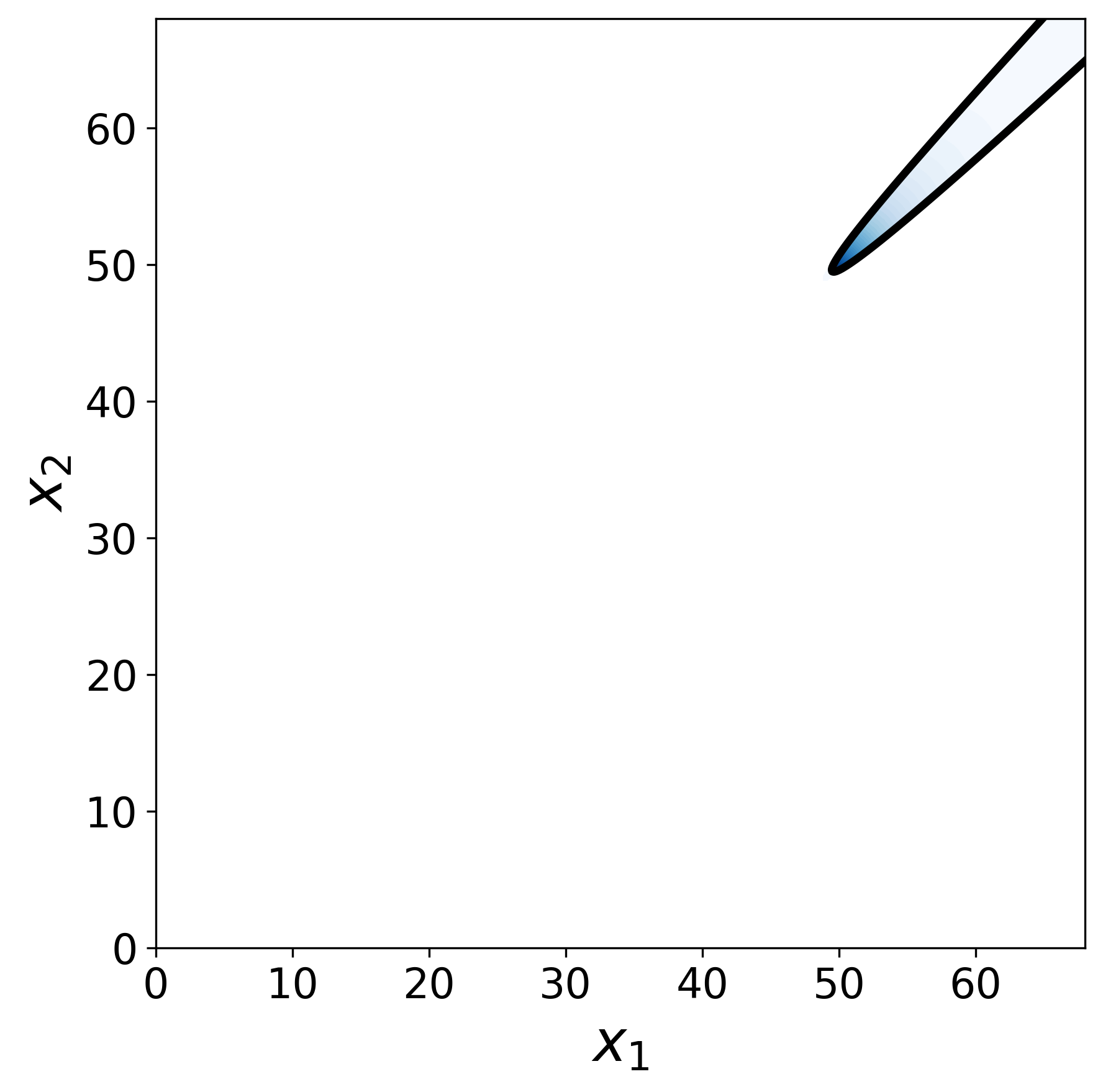}
        \caption{}
        \label{fig:two}
    \end{subfigure}\hfill
    \begin{subfigure}[t]{0.3\textwidth}
        \centering
        \includegraphics[width=\linewidth]{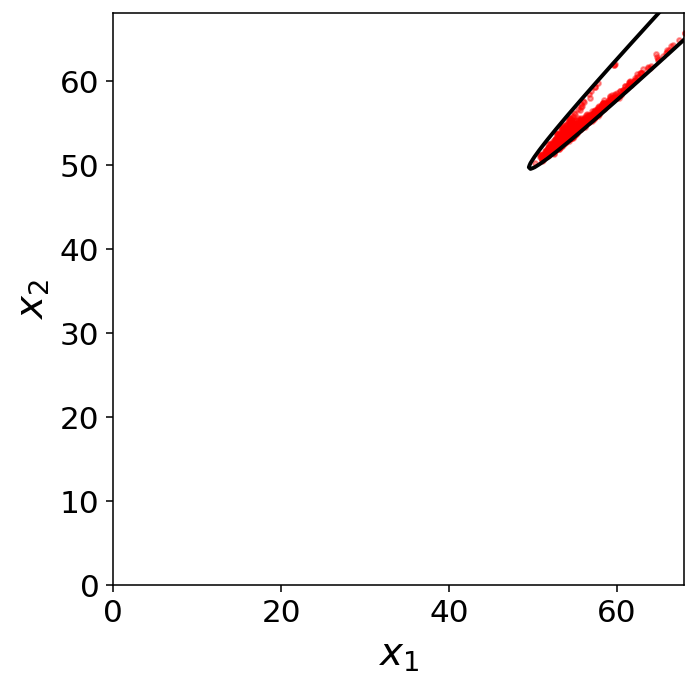}
        \caption{}
        \label{fig:three}
    \end{subfigure}

  \caption{\textbf{Example 3.} Visualization of the non-Gaussian reliability example with Gumbel marginals and Gaussian-copula dependence. 
(a) Joint correlated Gumbel input density (in red), together with the LSF \(S(\mathbf{x})=0\) (in black). 
(b) Approximate rare-event sampling target (in blue). 
(c) Final learned IS $N=1$ proposal samples (in red).}
\label{gumbel_copula_example}
\end{figure*}

Table~\ref{tab:lsf_K_N_results} highlights an important feature of FAMIS: the gain does not come only from increasing the sample size $K$ per iteration, but from matching the number of proposals to the geometry of the failure domain. For relatively simple or smoothly connected events, such as \(S_a\), \(S_b\), and \(S_d\), accurate estimates are obtained with only a small number of components, and the error decreases steadily as \(K\) increases. For the product-tail case \(S_c\), \(N=2\) yields large errors in the reported experiments, whereas \(N=4\) substantially reduces the RRMSE across all tested batch sizes. 
This confirms that proposal diversity is essential when the rare event contains multiple separated modes. The four-branch case \(S_e\) shows a similar behavior: additional components are most beneficial when $K$ is sufficiently large to support stable component updates. 
Overall, the results indicate that FAMIS can adapt effectively to a wide range of truncated-tail geometries, with the number of proposals influencing performance according to the complexity and multimodality of the failure region.

\subsection{Example 3: Non-Gaussian quadratic reliability example with correlated Gumbel marginals}
\label{subsec:gumbel_copula_example}

This example investigates the performance of the proposed method on a non-Gaussian reliability problem with a quadratic LSF. The input density follows a joint probability distribution constructed using a Gaussian copula with Gumbel marginal distributions written as
\begin{equation}
\widetilde{\pi}(\mathbf{x})
=
{\varphi}_{d_x}
\left(
\begin{bmatrix}
\Phi^{-1}\left(F(x_1)\right)\\
\Phi^{-1}\left(F(x_2)\right)\\
\vdots\\
\Phi^{-1}\left(F(x_{d_{x}})\right)
\end{bmatrix};
\boldsymbol{\Sigma}_c
\right)
\frac{
\prod_{i=1}^{d_x} f(x_i)
}{
\prod_{i=1}^{d_x}
\varphi\left(\Phi^{-1}\left(F(x_i)\right)\right)
},
\label{eq:gumbel_gaussian_copula_density}
\end{equation}
where \(\varphi_{d_x}(\cdot;\boldsymbol{\Sigma}_c)\) denotes the \(d_x\)-dimensional standard Gaussian density with correlation matrix \(\boldsymbol{\Sigma}_c\), while \(\varphi(\cdot)\) and \(\Phi(\cdot)\) denote the univariate standard Gaussian PDF and CDF, respectively. \(f(\cdot)\) and \(F(\cdot)\) are the PDF and CDF of the Gumbel marginal distribution of \(X_i\), respectively. In this example, all random variables $x_i$ have mean \(\mu_i=10\) and coefficient of variation \(0.40\). 
The limit state function is defined as
\begin{equation}
S(\mathbf{x})
=
\lambda
-
\frac{1}{\sqrt{d_x}}\sum_{i=1}^{d_x}x_i
+
2.5\left(x_1-\sum_{j=2}^{m}x_j\right)^2,
\label{eq:gumbel_lsf_general}
\end{equation}
where \(\lambda\) determines the rarity level and \(m\) specifies the number of nonlinear variables \cite{ESHRA2025111200}. 
The final failure probability estimates are obtained for dimensions $d_x\in \{2,3,40\}$, and two sample sizes, 
\(K \in\{100, 500\}\) generated by one proposal, $N=1$. In this case, $\alpha_1=1$ and the JSD term has no effect, so this example focuses on proposal adaptation rather than mixture repulsion. Table~\ref{tab:correlated_gumbel_results} reports the reference failure 
probability \(P_f\), the empirical mean estimate \(\mathbb{E}[\widehat{P}_f]\), 
the coefficient of variation \(\delta_{\widehat{P}_f}(\%)\), and RRMSE values.

For \(d_x=40\), the problem is 
more challenging due to the increased dimension and the nonlinear interaction 
among the first \(m\) variables. Nevertheless, FAMIS remains 
stable and produces estimates close to the reference probability 
\(P_f=4.60\times 10^{-6}\).  These results indicate that the learned proposal 
distribution successfully concentrates samples in the rare-event region despite the non-Gaussian and dependent structure of the input distribution. 
This is an important distinction from reliability methods \cite{wang2025structural, tong2021normal, wang2021hermite} that first transform the inputs into a standard Gaussian space and then perform IS in the transformed domain. Here, the normalizing flow learns a proposal directly over the original Gumbel-copula input variables.
Fig.~\ref{gumbel_copula_example} illustrates the joint probability distribution and the final learned proposal samples together with the limit state boundary \(S(\mathbf{x})=0\) for $d_x=2$, $\lambda=70$, and $m=2$. The learned proposal concentrate around the failure region.

\subsection{Example 4: High-dimensional benchmark with a highly nonlinear LSF}
\label{subsec:high_dim_nonlinear_benchmark}

We further assess the proposed framework on a high-dimensional benchmark problem defined in the standard normal space.  The LSF is given by \cite{PAPAKONSTANTINOU2023103485}
\begin{equation}
\begin{aligned}
S(\mathbf{x})
=&\; 4
-\frac{1}{\sqrt{d_x}}\sum_{i=1}^{d_x}x_i
+2.5\left(x_1-\sum_{i=2}^{\zeta_1}x_i\right)^2  \\
&+\xi_2\left(x_4-\sum_{i=5}^{\zeta_2}x_i\right)^4
+\xi_3\left(x_7-\sum_{i=8}^{\zeta_3}x_i\right)^8 .
\end{aligned}
\label{eq:high_dim_g8}
\end{equation}

To investigate the scalability of the proposed method, we set
\(\xi_2=\xi_3=1\), \(\zeta_1=3\), \(\zeta_2=6\), and \(\zeta_3=9\), while varying the input dimension \(d_x \in\{100,200,300\}\). This example is particularly challenging because the failure domain is governed by a combination of a high-dimensional linear term and strongly nonlinear quadratic, quartic, and eighth-order interactions. Table~\ref{high_dim_famis_rein_ss} compares the performance of FAMIS, REIN, and SS in terms of accuracy. Specifically, we compare the final estimate \(\mathbb{E}[\widehat{P}_f]\), \(\delta_{\widehat{P}_f}(\%)\), and MALE results based on $100$ independent simulation runs. REIN is implemented based on the general parameter choice recommendations reported in \cite{DASGUPTA2024109729}.

Table~\ref{high_dim_famis_rein_ss} shows that FAMIS, with  $N=2$ and $K=100$, remains accurate and stable across all tested dimensions, yielding the lowest MALE of the three methods. CWMH-SS produces mean estimates close to the reference values but exhibits larger variability, with $\delta_{\widehat{P}_f}$ between $23.39\%$ and $25.46\%$ and greater MALE between $0.197$ and $0.208$. Thus, FAMIS maintains competitive accuracy and stability as the dimension increases despite the strongly nonlinear failure geometry.

\begin{table*}[!t]
\centering
\caption{Example 4: Comparison of FAMIS, REIN, and SS performance for varying input dimension $d_x$. Bold values indicate the best performance for each metric.}
\label{high_dim_famis_rein_ss}
\renewcommand{\arraystretch}{1.55}
\resizebox{\linewidth}{!}{%
\begin{tabular}{c c c c c c c c c c c}
\toprule
\toprule
\multirow{2}{*}{$d_x$} 
& \multirow{2}{*}{$P_f$}
& \multicolumn{3}{c}{FAMIS}
& \multicolumn{3}{c}{REIN}
& \multicolumn{3}{c}{CWMH-SS} \\
\cmidrule(lr){3-5}\cmidrule(lr){6-8}\cmidrule(lr){9-11}
& 
& $\mathbb{E}[\widehat{P}_f]$
& $\delta_{\widehat{P}_f}(\%)$
& MALE
& $\mathbb{E}[\widehat{P}_f]$
& $\delta_{\widehat{P}_f}(\%)$
& MALE
& $\mathbb{E}[\widehat{P}_f]$
& $\delta_{\widehat{P}_f}(\%)$
& MALE \\
\midrule

$100$
& $3.43\times 10^{-7}$
& $\mathbf{3.43\times 10^{-7}}$
& $\mathbf{6.01}$
& $\mathbf{0.060}$
& $3.45\times 10^{-7}$
& $12.00$
& $0.120$
& $3.55\times 10^{-7}$
& $23.39$
& $0.197$ \\

\midrule

$200$
& $3.55\times 10^{-7}$
& $\mathbf{3.50\times 10^{-7}}$
& $\mathbf{10.32}$
& $\mathbf{0.101}$
& $3.66\times 10^{-7}$
& $15.50$
& $0.158$
& ${3.67\times 10^{-7}}$
& $25.03$
& $0.201$ \\

\midrule

$300$
& $3.71\times 10^{-7}$
& $\mathbf{3.59\times 10^{-7}}$
& $13.00$
& $\mathbf{0.120}$
& $3.57\times 10^{-7}$
& $\mathbf{12.80}$
& $0.134$
& $3.52\times 10^{-7}$
& $25.46$
& $0.208$ \\

\bottomrule
\bottomrule
\end{tabular}
}
\end{table*}

\begin{table*}[!htbp]
\centering
\scriptsize
\caption{Example 5: Comparison of different methods for varying high dimensions $d_x$ under the funnel distribution. Bold values indicate the best performance for each metric.}
\label{neal_funnel_results}
\renewcommand{\arraystretch}{1.45}
\setlength{\tabcolsep}{3.5pt}
\resizebox{\linewidth}{!}{%
\begin{tabular}{c c c c c c c c c c c c c c}
\toprule
\toprule
\multirow{2}{*}{$d_x$}
& \multirow{2}{*}{$P_f$}
& \multicolumn{3}{c}{FAMIS}
& \multicolumn{3}{c}{REIN}
& \multicolumn{3}{c}{ASTPA-HMCMC}
& \multicolumn{3}{c}{CWMH-SS} \\
\cmidrule(lr){3-5}\cmidrule(lr){6-8}\cmidrule(lr){9-11}\cmidrule(lr){12-14}
&
& $\mathbb{E}[\widehat{P}_f]$
& $\delta_{\widehat{P}_f}(\%)$
& MALE
& $\mathbb{E}[\widehat{P}_f]$
& $\delta_{\widehat{P}_f}(\%)$
& MALE
& $\mathbb{E}[\widehat{P}_f]$
& $\delta_{\widehat{P}_f}(\%)$
& MALE
& $\mathbb{E}[\widehat{P}_f]$
& $\delta_{\widehat{P}_f}(\%)$
& MALE \\
\midrule

$2$
& $3.10\times 10^{-5}$
& $\mathbf{3.10\times 10^{-5}}$
& $\mathbf{0.52}$
& $\mathbf{0.005}$
& $3.11\times 10^{-5}$
& $0.87$
& $0.007$
& $3.09\times 10^{-5}$
& $9.00$
& $0.072$
& $3.12\times 10^{-5}$
& $10.84$
& $0.089$ \\

\midrule

$31$
& $1.87\times 10^{-5}$
& $\mathbf{1.87\times 10^{-5}}$
& $3.99$
& $\mathbf{0.042}$
& $1.52\times 10^{-5}$
& $\mathbf{1.90}$
& $1.362$
& $1.83\times 10^{-5}$
& $10.00$
& $0.082$
& $1.92\times 10^{-5}$
& $13.03$
& $0.120$ \\

\midrule

$51$
& $1.37\times 10^{-5}$
& $\mathbf{1.36\times 10^{-5}}$
& $7.36$
& $\mathbf{0.075}$
& $3.03\times 10^{-6}$
& $\mathbf{3.32}$
& $6.054$
& $1.34\times 10^{-5}$
& $15.00$
& $0.122$
& $1.32\times 10^{-5}$
& $12.32$
& $0.131$ \\

\midrule

$101$
& $6.85\times 10^{-6}$
& $\mathbf{6.74\times 10^{-6}}$
& $\mathbf{3.07}$
& $\mathbf{0.048}$
& -- 
& -- 
& -- 
& $6.55\times 10^{-6}$
& $16.00$
& $0.134$
& $6.98\times 10^{-6}$
& $14.65$
& $0.220$ \\

\bottomrule
\bottomrule
\end{tabular}
}
\end{table*}
\subsection{Example 5: High-dimensional hyperspherical LSF With Neal's funnel distribution}
\label{funnel_hypersphere}

We next evaluate the proposed framework on a challenging non-Gaussian reliability problem involving a high-dimensional Neal's funnel distribution \cite{ESHRA2025111200, CHEN2022}.  The joint density is defined as
\begin{equation}
\widetilde{\pi}(\mathbf{x})
=
\left[
\prod_{i=1}^{d_x-1}
\mathcal{N}\left(x_i \mid 0,\exp(x_{d_x})\right)
\right]
\mathcal{N}\left(x_{d_x} \mid 0,1\right),
\label{eq:funnel_density}
\end{equation}
where the last coordinate \(x_{d_x}\) controls the conditional scale of the first \(d_x-1\) variables. This distribution is highly non-Gaussian and exhibits strong nonlinear dependence through the variance term \(\exp(x_{d_x})\), making it a difficult target for rare-event simulation.
The reliability problem is defined through the following hyperspherical LSF:
\begin{equation}
S(\mathbf{x})
=
\sum_{i=1}^{d_x-1}x_i^2
+
(x_{d_x}+6)^2
-
r^2,
\label{eq:funnel_lsf}
\end{equation}
where \(r\) is a threshold parameter controlling the rarity of the failure event. The failure domain $\mathcal{F}$ corresponds to the interior of a \(d_x\)-dimensional hypersphere centered at
\((0,\ldots,0,-6)\) with radius \(r\).  The target distribution has a funnel-shaped geometry: when \(x_{d_x}\) is small, the conditional variance of \(x_1,\ldots,x_{d_x-1}\) is small, whereas large values of \(x_{d_x}\) induce much wider conditional distributions. At the same time, the rare-event region is defined by a hyperspherical constraint shifted toward the lower tail of \(x_{d_x}\). As a result, efficient sampling requires the proposal to adapt both to the nonlinear dependence structure of the input distribution and to the localized failure region.

\begin{figure*}[!b]
    \centering
    \begin{subfigure}[t]{0.3\textwidth}
        \centering
        \includegraphics[width=\linewidth]{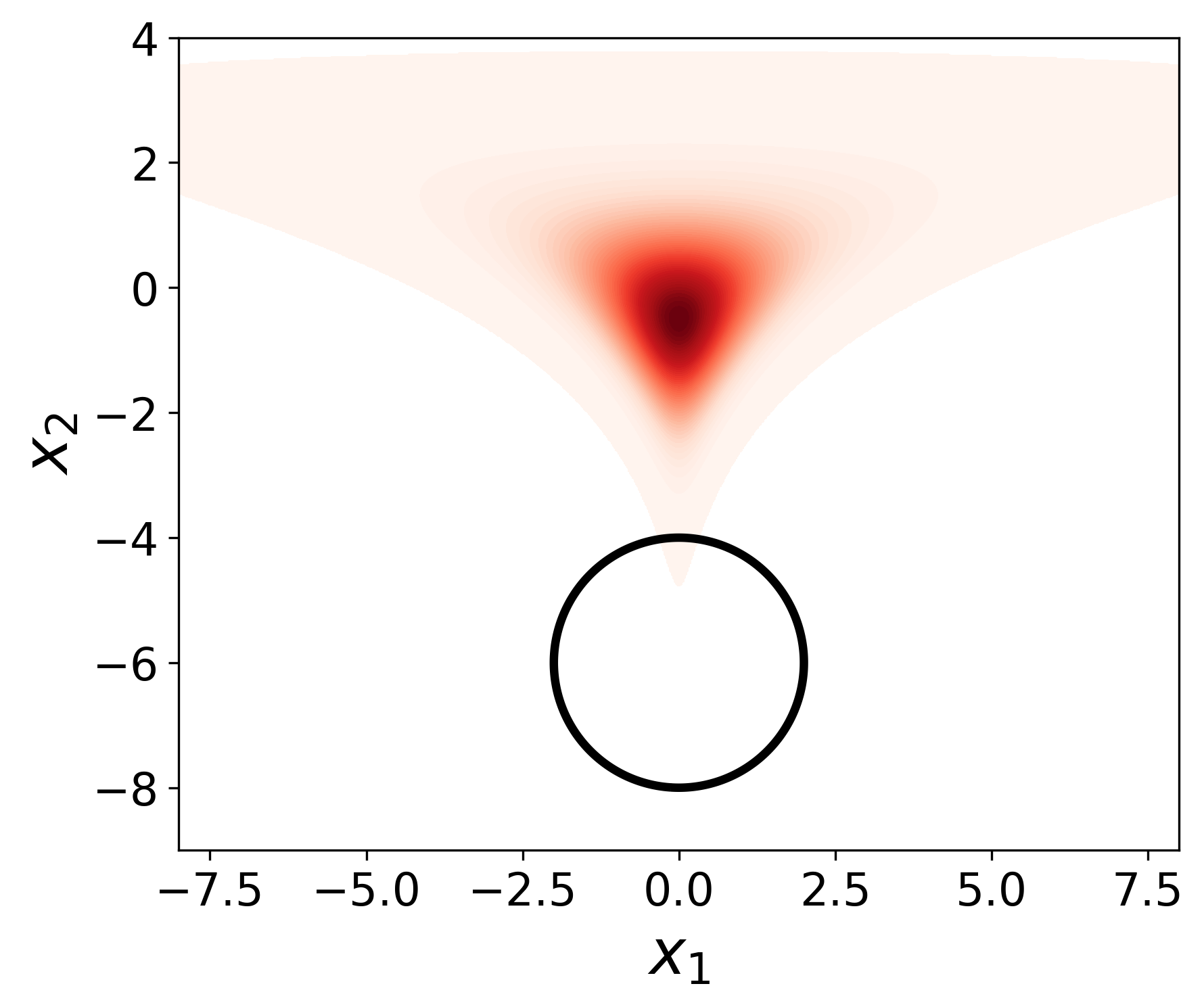}
        \caption{}
        \label{fig:one}
    \end{subfigure}\hfill
    \begin{subfigure}[t]{0.3\textwidth}
        \centering
        \includegraphics[width=\linewidth]{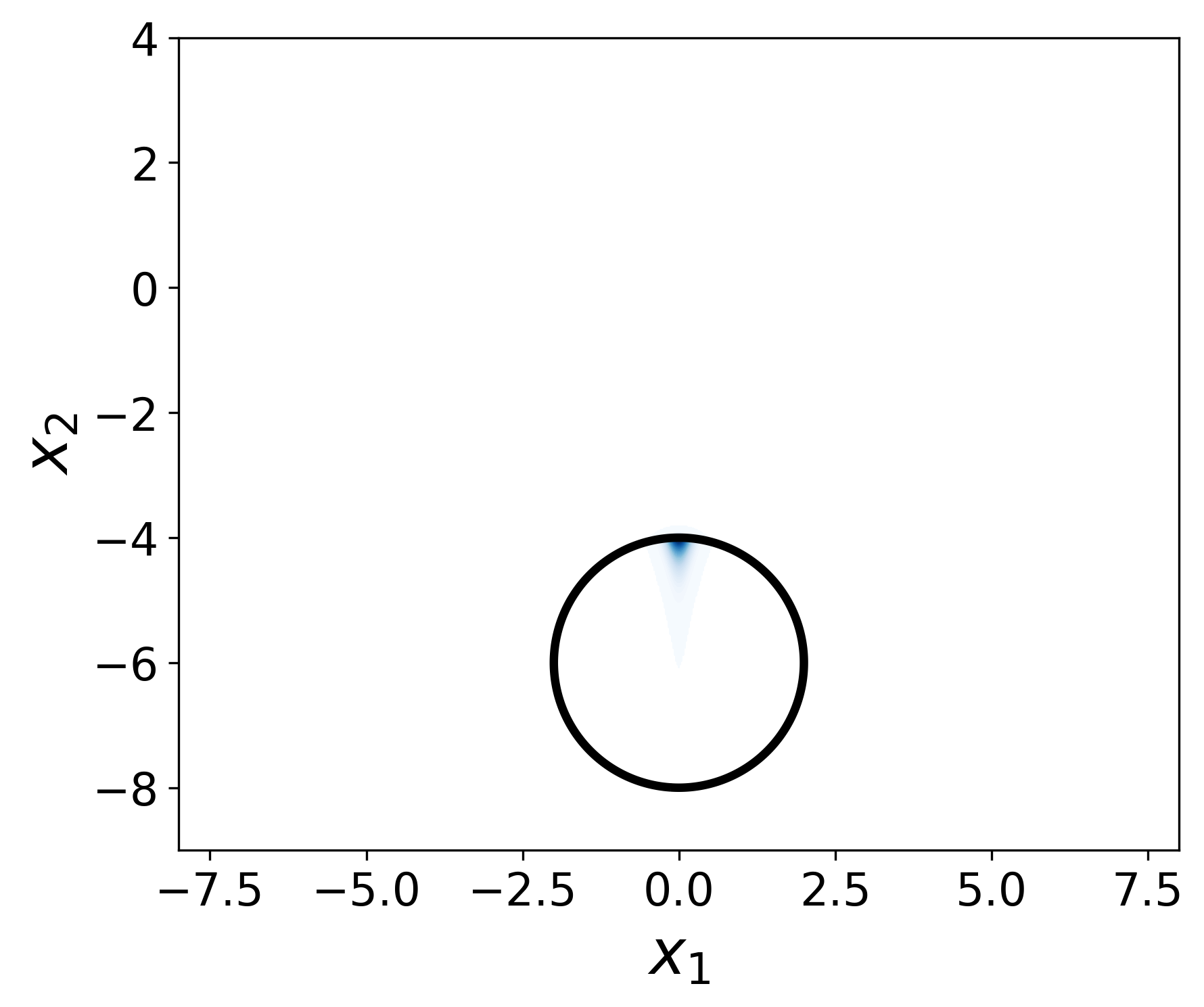}
        \caption{}
        \label{fig:two}
    \end{subfigure}\hfill
    \begin{subfigure}[t]{0.3\textwidth}
        \centering
        \includegraphics[width=\linewidth]{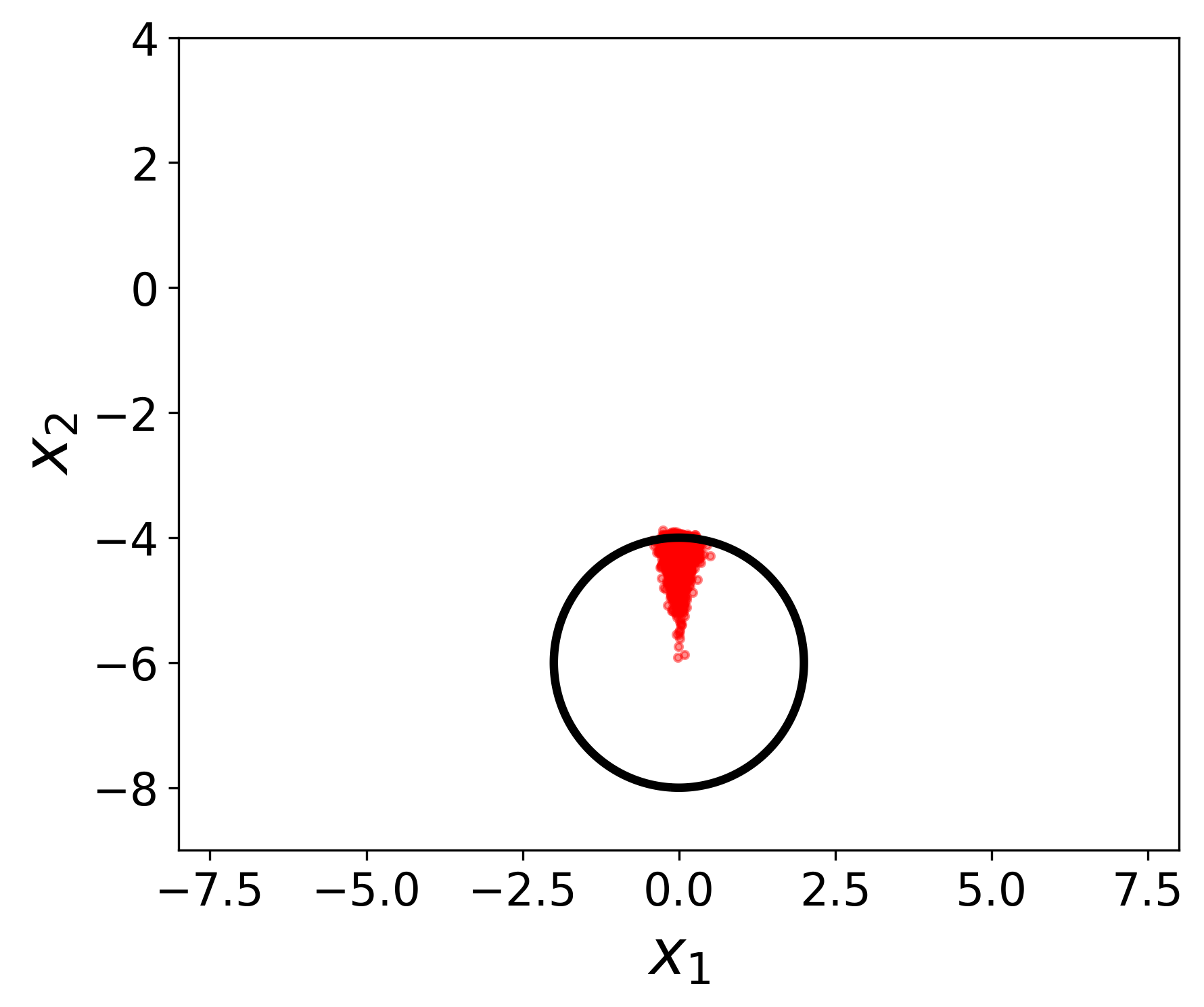}
        \caption{}
        \label{fig:three}
    \end{subfigure}

  \caption{\textbf{Example 5.} Visualization of the Neal's funnel reliability example. 
(a) Neal's funnel target density (red), together with the LSF \(S(\mathbf{x})=0\) (in black). 
(b) Approximate rare-event sampling target (in blue)
(c) Final learned IS $N=1$ proposal samples (in red) for visualization.}
\label{funne_example}
\end{figure*}

For the quantitative results, FAMIS uses $K=100$ and $N=2$ flow proposals learned directly in the original funnel input space. 
The numerical results are reported in Table~\ref{neal_funnel_results} for dimensions up to $d_x=101$. The performance of ASTPA-HMCMC is shown using the available results reported in \cite{ESHRA2025111200}. Fig.~\ref{funne_example} illustrates the funnel input distribution $\widetilde{\pi}(\boldsymbol{x})$ and the hyperspherical failure boundary. It also illustrates the approximate sampling target for $d_x=2$ and $r=2$ using $N=1$ to visualize the learned geometry rather than the effect of repulsion. 

Table~\ref{neal_funnel_results} shows that the main difficulty in this example is not merely reducing estimator variance, but placing probability mass in the correct tail of the funnel distribution. This is evident from the comparison between variability and logarithmic error: some methods achieve small coefficients of variation while still producing large MALE values, indicating stable but biased exploration of the wrong part of the rare-event geometry. In contrast, FAMIS maintains close agreement with the reference probabilities across all dimensions, with MALE values remaining small even as \(d_x\) increases. The improvement is particularly clear in the high-dimensional cases, where REIN deteriorates sharply. ASTPA-HMCMC provides comparable estimates, but its error increases with dimension, whereas FAMIS preserves both accuracy and stability. Additionally, at \(d_x=51\) and \(d_x=101\), CWMH-SS becomes competitive with ASTPA-HMCMC in terms of estimator variability, but requires substantially more LSF evaluations (i.e., $2.30\times10^5$ and $2.75\times10^5$, respectively). These results suggest that the proposed mixture of repulsive flow proposals is geometry-aware and reduces variance. It can locate the relevant failure region under the nonlinear dependence and unconstrained scale  induced by Neal's funnel.

\subsection{Computational Complexity}
\label{subsec:computational_complexity}

The computational cost of FAMIS is governed by the number of samples \(K\) drawn at each adaptive iteration, or equivalently by the number of LSF evaluations. Since FAMIS uses a shared minibatch of \(K\) samples from the defensive mixture ${{\Psi}}^{(t)}$ at every iteration $t$, the training stage requires \(KT\) LSF calls. If \(K_{\mathrm{est}}{\sim}q_{\alpha}^{(T)}\) samples at last iteration $T$  are used to compute the final estimate \(\widehat{P}_f\), the total number of LSF evaluations is
\begin{equation}
N_{\mathrm{call}} = KT + K_{\mathrm{est}} .
\end{equation}
Thus, the LSF evaluation cost scales linearly as \(N_{\mathrm{call}}=\mathcal{O}(KT+K_{\mathrm{est}})\). This fixed computational budget is known a priori and is independent of problem complexity. 

\begin{figure}[!h]
    \centering  \includegraphics[width=0.55\textwidth]{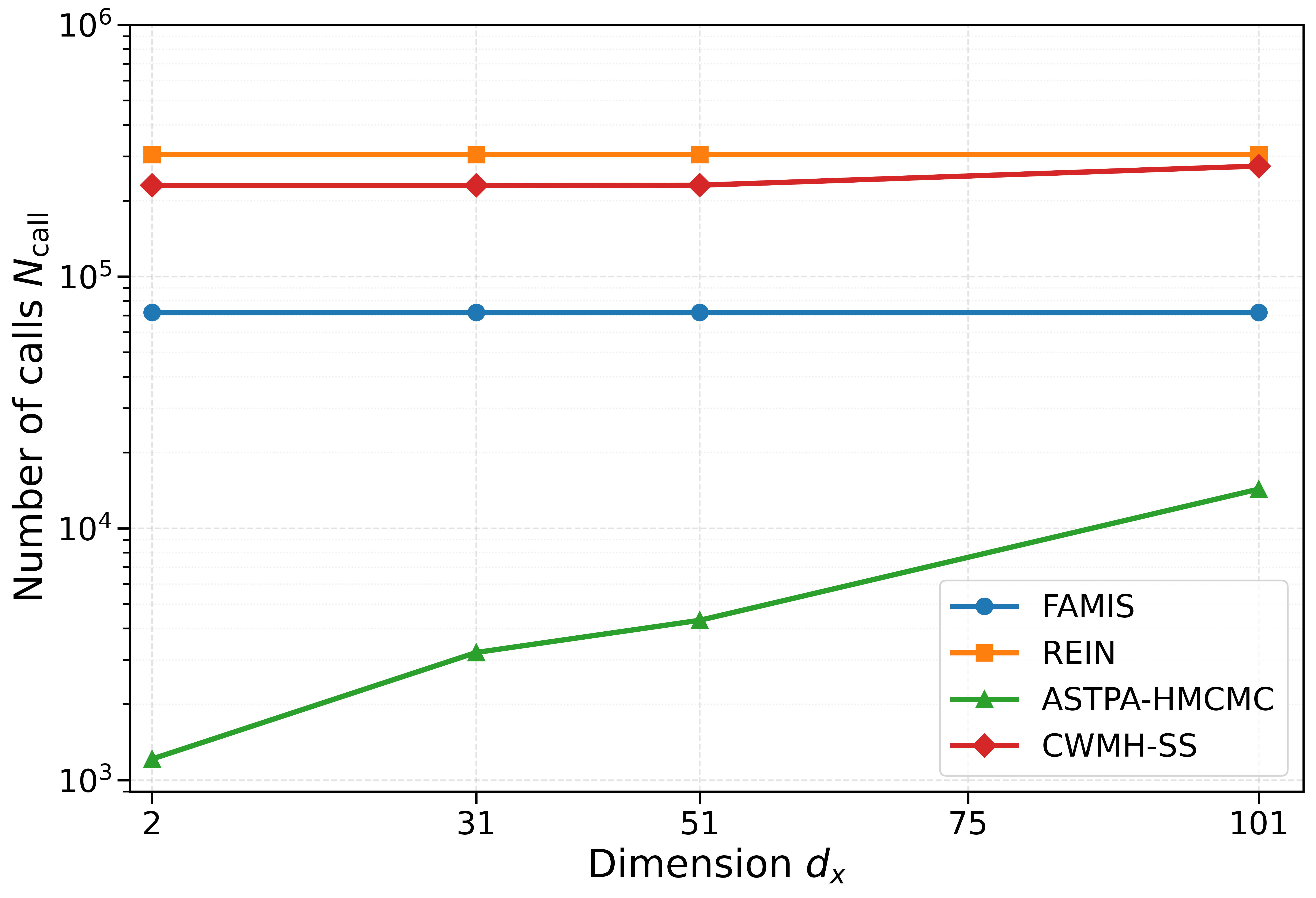}
    
    \caption{LSF evaluation cost $N_{\mathrm{call}}$ versus dimension
$d_x$ for the funnel benchmark in Example~5.}
    \label{computational_complexity}
\end{figure}
Figure~\ref{computational_complexity} compares $N_{\mathrm{call}}$ for FAMIS, REIN, ASTPA, and CWMH-SS on the funnel distribution benchmark described in Example~5. FAMIS uses a fixed budget of \(N_{\text{call}}=72\times 10^3\),
notably fewer than CWMH-SS and REIN, which requires \(N_{\text{call}}=3.05 \times 10^5\).
Although ASTPA-HMCMC requires fewer LSF evaluations, Table~\ref{neal_funnel_results} shows a clear tradeoff between  accuracy and cost. ASTPA-HMCMC attains MALE values up to an order of magnitude larger than those of FAMIS.
In contrast, FAMIS maintains $\mathrm{MALE}\leq 0.075$ for all $d_x$ and remains accurate at $d_x=101$, where REIN fails. These results indicate that the additional FAMIS evaluations enable more reliable and robust exploration of the rare-event region and ensure estimator stability and accuracy.

\section{Conclusion}
\label{conc}

This paper presented FAMIS, an adaptive flow-based multiple importance sampling framework for rare-event probability estimation. FAMIS learns flexible failure-focused proposals directly from sequential limit state evaluations. The proposed method does not require presampled failures, prior knowledge of the number, location, or geometry of failure modes, or gradients of the limit state function. By combining shared normalizing flow architecture, adaptive mixture weighting, defensive exploration, and repulsive component learning, the method is designed to maintain coverage of complex and high-dimensional rare events while maintaining unbiased estimation via importance weights. Through a set of multimodal, non-Gaussian, and strongly nonlinear reliability problems, FAMIS consistently produced accurate and stable estimates of very small failure probabilities, with low estimator variance and strong performance relative to the considered benchmark methods. These results demonstrate the potential of FAMIS as a flexible and computationally efficient framework for challenging reliability analysis of complex systems.


\section*{Acknowledgements}
The authors gratefully acknowledge the support from the EPSRC Centre for Doctoral Training in Mathematical Modeling, Analysis and Computation (MAC-MIGS) funded by the UK Engineering and Physical Sciences Research Council (grant EP/S023291/1) and ARL/ARO under grant W911NF-22-1-0235.
\bibliographystyle{elsarticle-num}  
\bibliography{refs}

@article{hesterberg1995weighted,
  title={Weighted average importance sampling and defensive mixture distributions},
  author={Hesterberg, Tim},
  journal={Technometrics},
  volume={37},
  number={2},
  pages={185--194},
  year={1995},
  publisher={Taylor \& Francis}
}

@article{cappe2008adaptive,
  title={Adaptive importance sampling in general mixture classes},
  author={Capp{\'e}, Olivier and Douc, Randal and Guillin, Arnaud and Marin, Jean-Michel and Robert, Christian P},
  journal={Statistics and Computing},
  volume={18},
  number={4},
  pages={447--459},
  year={2008},
  publisher={Springer}
}

@article{elvira2019generalized,
author = {V{\'i}ctor Elvira and Luca Martino and David Luengo and M{\'o}nica F. Bugallo},
title = {{Generalized Multiple Importance Sampling}},
volume = {34},
journal = {Statistical Science},
number = {1},
publisher = {Institute of Mathematical Statistics},
pages = {129 -- 155},
year = {2019},
}

@article{DASGUPTA2024109729,
title = {REIN: Reliability Estimation via Importance sampling with Normalizing flows},
journal = {Reliability Engineering \& System Safety},
volume = {242},
pages = {109729},
year = {2024},
author = {Agnimitra Dasgupta and Erik A. Johnson},
}

@misc{gibson2023flow,
      title={A Flow-Based Generative Model for Rare-Event Simulation}, 
      author={Lachlan Gibson and Marcus Hoerger and Dirk Kroese},
      year={2023},
      eprint={2305.07863},
      archivePrefix={arXiv},
      primaryClass={stat.ML},
}

@misc{gao2024nofis,
      title={Rare Event Probability Learning by Normalizing Flows}, 
      author={Zhenggqi Gao and Dinghuai Zhang and Luca Daniel and Duane S. Boning},
      year={2023},
      eprint={2310.19167},
      archivePrefix={arXiv},
      primaryClass={cs.LG},
}

@article{asghar2024flowres,
  title={Efficient rare event sampling with unsupervised normalizing flows},
  author={Asghar, Solomon and Pei, Qing-Xiang and Volpe, Giorgio and Ni, Ran},
  journal={Nature Machine Intelligence},
  volume={6},
  number={11},
  pages={1370--1381},
  year={2024},
  publisher={Nature Publishing Group UK London}
}

@inproceedings{dawson2025rare,
  title={Rare event modeling with self-regularized normalizing flows: what can we learn from a single failure?},
  author={Dawson, Charles and Tran, Van and Li, Max and Fan, Chuchu},
  booktitle={International Conference on Learning Representations},
  volume={2025},
  pages={28763--28784},
  year={2025}
}

@article{cadini2017estimation,
  title={Estimation of rare event probabilities in power transmission networks subject to cascading failures},
  author={Cadini, Francesco and Agliardi, Gian Luca and Zio, Enrico},
  journal={Reliability Engineering \& System Safety},
  volume={158},
  pages={9--20},
  year={2017},
  publisher={Elsevier}
}

@article{wei2019structural,
  title={Structural reliability and reliability sensitivity analysis of extremely rare failure events by combining sampling and surrogate model methods},
  author={Wei, Pengfei and Tang, Chenghu and Yang, Yuting},
  journal={Proceedings of the Institution of Mechanical Engineers, Part O: Journal of risk and reliability},
  volume={233},
  number={6},
  pages={943--957},
  year={2019},
  publisher={SAGE Publications Sage UK: London, England}
}

@book{morio2015estimation,
  title={Estimation of rare event probabilities in complex aerospace and other systems: a practical approach},
  author={Morio, J{\'e}r{\^o}me and Balesdent, Mathieu},
  year={2015},
  publisher={Woodhead publishing}
}

@INPROCEEDINGS{10711532BAI,
  author={Bai, Ruoxuan and Yang, Jingxuan and Gong, Weiduo and Zhang, Yi and Lu, Qiujing and Feng, Shuo},
  booktitle={2024 IEEE 20th International Conference on Automation Science and Engineering (CASE)}, 
  title={Accurately Predicting Probabilities of Safety-Critical Rare Events for Intelligent Systems}, 
  year={2024},
  volume={},
  number={},
  pages={3243-3249},
  }

@article{keshtegar2017hybrid,
  title={A hybrid relaxed first-order reliability method for efficient structural reliability analysis},
  author={Keshtegar, Behrooz and Meng, Zeng},
  journal={Structural Safety},
  volume={66},
  pages={84--93},
  year={2017},
  publisher={Elsevier}
}

@article{gong2017first,
  title={First-order reliability method-based system reliability analyses of corroding pipelines considering multiple defects and failure modes},
  author={Gong, Changqing and Zhou, Wenxing},
  journal={Structure and Infrastructure Engineering},
  volume={13},
  number={11},
  pages={1451--1461},
  year={2017},
  publisher={Taylor \& Francis}
}

@article{lim2016post,
  title={Post optimization for accurate and efficient reliability-based design optimization using second-order reliability method based on importance sampling and its stochastic sensitivity analysis},
  author={Lim, Jongmin and Lee, Byungchai and Lee, Ikjin},
  journal={International Journal for Numerical Methods in Engineering},
  volume={107},
  number={2},
  pages={93--108},
  year={2016},
  publisher={Wiley Online Library}
}

@article{zhao1999general,
  title={A general procedure for first/second-order reliability method (FORM/SORM)},
  author={Zhao, Yan-Gang and Ono, Tetsuro},
  journal={Structural safety},
  volume={21},
  number={2},
  pages={95--112},
  year={1999},
  publisher={Elsevier}
}

@article{lucia2763,
author = {Lucia Faravelli },
title = {Response‐Surface Approach for Reliability Analysis},
journal = {Journal of Engineering Mechanics},
volume = {115},
number = {12},
pages = {2763-2781},
year = {1989},
}

@article{PCK0000870,
author = {R. Schöbi  and B. Sudret  and S. Marelli },
title = {Rare Event Estimation Using Polynomial-Chaos Kriging},
journal = {ASCE-ASME Journal of Risk and Uncertainty in Engineering Systems, Part A: Civil Engineering},
volume = {3},
pages={D4016002},
number = {2},
year = {2017},
 }

@article{schobi2015polynomial,
  title={Polynomial-chaos-based Kriging},
  author={Schobi, Roland and Sudret, Bruno and Wiart, Joe},
  journal={International Journal for Uncertainty Quantification},
  volume={5},
  number={2},
  year={2015},
  publisher={Begel House Inc.}
}

@article{HURTADO2004271,
title = {An examination of methods for approximating implicit limit state functions from the viewpoint of statistical learning theory},
journal = {Structural Safety},
volume = {26},
number = {3},
pages = {271-293},
year = {2004},
author = {Jorge E. Hurtado},
}

@article{zhang2022moving,
  title={Moving-zone renewal strategy combining adaptive Kriging and truncated importance sampling for rare event analysis},
  author={Zhang, Hong and Song, Lu-Kai and Bai, Guang-Chen},
  journal={Structural and Multidisciplinary Optimization},
  volume={65},
  number={10},
  pages={285},
  year={2022},
  publisher={Springer}
}

@article{bichon2008efficient,
  title={Efficient global reliability analysis for nonlinear implicit performance functions},
  author={Bichon, Barron J and Eldred, Michael S and Swiler, Laura Painton and Mahadevan, Sandaran and McFarland, John M},
  journal={AIAA journal},
  volume={46},
  number={10},
  pages={2459--2468},
  year={2008}
}

@article{BAO2021107778,
title = {Adaptive subset searching-based deep neural network method for structural reliability analysis},
journal = {Reliability Engineering \& System Safety},
volume = {213},
pages = {107778},
year = {2021},
author = {Yuequan Bao and Zhengliang Xiang and Hui Li},
}

@article{ELVIRA201777,
title = {Improving population Monte Carlo: Alternative weighting and resampling schemes},
journal = {Signal Processing},
volume = {131},
pages = {77-91},
year = {2017},
author = {Víctor Elvira and Luca Martino and David Luengo and Mónica F. Bugallo},
}

@article{elvira2022optimized,
  title={Optimized population monte carlo},
  author={Elvira, V{\'\i}ctor and Chouzenoux, Emilie},
  journal={IEEE Transactions on Signal Processing},
  volume={70},
  pages={2489--2501},
  year={2022},
  publisher={IEEE}
}

@article{cappe2004population,
  title={Population monte carlo},
  author={Capp{\'e}, Olivier and Guillin, Arnaud and Marin, Jean-Michel and Robert, Christian P},
  journal={Journal of Computational and Graphical Statistics},
  volume={13},
  number={4},
  pages={907--929},
  year={2004},
  publisher={Taylor \& Francis}
}

@inproceedings{martino2014adaptive,
  title={An adaptive population importance sampler},
  author={Martino, Luca and Elvira, Victor and Luengo, David and Corander, Jukka},
  booktitle={2014 IEEE International Conference on Acoustics, Speech and Signal Processing (ICASSP)},
  pages={8038--8042},
  year={2014},
  organization={IEEE}
}

@article{elvira2021advances,
  title={Advances in importance sampling},
  author={Elvira, V{\'\i}ctor and Martino, Luca},
  journal={arXiv preprint arXiv:2102.05407},
  year={2021}
}

@article{AU2001263,
title = {Estimation of small failure probabilities in high dimensions by subset simulation},
journal = {Probabilistic Engineering Mechanics},
volume = {16},
number = {4},
pages = {263-277},
year = {2001},
author = {Siu-Kui Au and James L. Beck},
}

@article{PRADLWARTER2007208,
title = {Application of line sampling simulation method to reliability benchmark problems},
journal = {Structural Safety},
volume = {29},
number = {3},
pages = {208-221},
year = {2007},
author = {H.J. Pradlwarter and G.I. Schuëller and P.S. Koutsourelakis and D.C. Charmpis},
}

@article{grooteman2011adaptive,
  title={An adaptive directional importance sampling method for structural reliability},
  author={Grooteman, Frank},
  journal={Probabilistic Engineering Mechanics},
  volume={26},
  number={2},
  pages={134--141},
  year={2011},
  publisher={Elsevier}
}

@article{guo2020active,
  title={An active learning Kriging model combined with directional importance sampling method for efficient reliability analysis},
  author={Guo, Qing and Liu, Yongshou and Chen, Bingqian and Zhao, Yuzhen},
  journal={Probabilistic Engineering Mechanics},
  volume={60},
  pages={103054},
  year={2020},
  publisher={Elsevier}
}

@article{XIAO2020106852,
title = {Reliability analysis with stratified importance sampling based on adaptive Kriging},
journal = {Reliability Engineering \& System Safety},
volume = {197},
pages = {106852},
year = {2020},
author = {Sinan Xiao and Sergey Oladyshkin and Wolfgang Nowak},
}

@article{22M1522838,
author = {Chennetier, Guillaume and Chraibi, Hassane and Dutfoy, Anne and Garnier, Josselin},
title = {Adaptive Importance Sampling Based on Fault Tree Analysis for Piecewise Deterministic Markov Process},
journal = {SIAM/ASA Journal on Uncertainty Quantification},
volume = {12},
number = {1},
pages = {128-156},
year = {2024},
}

@article{CHENG2023102291,
title = {Rare event estimation with sequential directional importance sampling},
journal = {Structural Safety},
volume = {100},
pages = {102291},
year = {2023},
author = {Kai Cheng and Iason Papaioannou and Zhenzhou Lu and Xiaobo Zhang and Yanping Wang},
}

@article{22M1524758,
author = {Tong, Shanyin and Stadler, Georg},
title = {Large Deviation Theory-based Adaptive Importance Sampling for Rare Events in High Dimensions},
journal = {SIAM/ASA Journal on Uncertainty Quantification},
volume = {11},
number = {3},
pages = {788-813},
year = {2023},
}

@article{XIAN2024102393,
title = {Relaxation-based importance sampling for structural reliability analysis},
journal = {Structural Safety},
volume = {106},
pages = {102393},
year = {2024},
author = {Jianhua Xian and Ziqi Wang},
}

@article{uribe2021cross,
author = {Uribe, Felipe and Papaioannou, Iason and Marzouk, Youssef M. and Straub, Daniel},
title = {Cross-Entropy-Based Importance Sampling with Failure-Informed Dimension Reduction for Rare Event Simulation},
journal = {SIAM/ASA Journal on Uncertainty Quantification},
volume = {9},
number = {2},
pages = {818-847},
year = {2021},
}

@article{geyer2019cross,
title = {Cross entropy-based importance sampling using Gaussian densities revisited},
journal = {Structural Safety},
volume = {76},
pages = {15-27},
year = {2019},
author = {Sebastian Geyer and Iason Papaioannou and Daniel Straub},
}

@article{PAPAIOANNOU201666,
title = {Sequential importance sampling for structural reliability analysis},
journal = {Structural Safety},
volume = {62},
pages = {66-75},
year = {2016},
author = {Iason Papaioannou and Costas Papadimitriou and Daniel Straub},
}

@INPROCEEDINGS{Elvira8682284,
  author={Elvira, V{\'\i}ctor and Chouzenoux, {\'{E}}milie},
  booktitle={2019 IEEE International Conference on Acoustics, Speech and Signal Processing (ICASSP)}, 
  title={Langevin-based Strategy for Efficient Proposal Adaptation in Population Monte Carlo}, 
  year={2019},
  volume={},
  number={},
  pages={5077-5081},
  }

@ARTICLE{Bugallo7974876,
  author={Bugallo, Monica F. and Elvira, Victor and Martino, Luca and Luengo, David and Miguez, Joaquin and Djuric, Petar M.},
  journal={IEEE Signal Processing Magazine}, 
  title={Adaptive Importance Sampling: The past, the present, and the future}, 
  year={2017},
  volume={34},
  number={4},
  pages={60-79},
  }

@inproceedings{rezende2015variational,
  title={Variational Inference with Normalizing Flows},
  author={Rezende, Danilo and Mohamed, Shakir},
  booktitle={Proceedings of the 32nd International Conference on Machine Learning},
  pages={1530--1538},
  year={2015}
}

@article{kobyzev2020normalizing,
  title={Normalizing Flows: An Introduction and Review of Current Methods},
  author={Kobyzev, Ivan and Prince, Simon J. D. and Brubaker, Marcus A.},
  journal={IEEE Transactions on Pattern Analysis and Machine Intelligence},
  volume={43},
  number={11},
  pages={3964--3979},
  year={2021}
}

@article{papamakarios2021normalizing,
  title={Normalizing Flows for Probabilistic Modeling and Inference},
  author={Papamakarios, George and Nalisnick, Eric and Rezende, Danilo Jimenez and Mohamed, Shakir and Lakshminarayanan, Balaji},
  journal={Journal of Machine Learning Research},
  volume={22},
  number={57},
  pages={1--64},
  year={2021}
}

@article{dinh2014nice,
  title={NICE: Non-linear Independent Components Estimation},
  author={Dinh, Laurent and Krueger, David and Bengio, Yoshua},
  journal={arXiv preprint arXiv:1410.8516},
  year={2014}
}

@article{dinh2016realnvp,
  title={Density Estimation using Real NVP},
  author={Dinh, Laurent and Sohl-Dickstein, Jascha and Bengio, Samy},
  journal={arXiv preprint arXiv:1605.08803},
  year={2016}
}

@inproceedings{kingma2016iaf,
  title={Improving Variational Inference with Inverse Autoregressive Flow},
  author={Kingma, Diederik P. and Salimans, Tim and Jozefowicz, Rafal and Chen, Xi and Sutskever, Ilya and Welling, Max},
  booktitle={Advances in Neural Information Processing Systems},
  year={2016}
}

@inproceedings{papamakarios2017maf,
  title={Masked Autoregressive Flow for Density Estimation},
  author={Papamakarios, George and Pavlakou, Theo and Murray, Iain},
  booktitle={Advances in Neural Information Processing Systems},
  year={2017}
}

@inproceedings{durkan2019neural,
  title={Neural Spline Flows},
  author={Durkan, Conor and Bekasov, Artur and Murray, Iain and Papamakarios, George},
  booktitle={Advances in Neural Information Processing Systems},
  year={2019}
}

@article{ESHRA2025111200,
title = {A direct importance sampling-based framework for rare event uncertainty quantification in non-Gaussian spaces},
journal = {Reliability Engineering \& System Safety},
volume = {264},
pages = {111200},
year = {2025},
author = {Elsayed Eshra and Konstantinos G. Papakonstantinou and Hamed Nikbakht},
}

@article{douc2008convergence,
  title={Convergence of adaptive mixtures of importance sampling schemes},
  author={Douc, Randal and Guillin, A and Marin, Jean-Michel and Robert, CP},
  journal={Statistics and Computing},
  volume={18},
  pages={447--459},
  year={2008}
}

@article{CHEN2022,
title = {Riemannian Manifold Hamiltonian Monte Carlo based subset simulation for reliability analysis in non-Gaussian space},
journal = {Structural Safety},
volume = {94},
pages = {102134},
year = {2022},
author = {Weiming Chen and Ziqi Wang and Marco Broccardo and Junho Song},
}

@article{PAPAKONSTANTINOU2023103485,
title = {Hamiltonian MCMC methods for estimating rare events probabilities in high-dimensional problems},
journal = {Probabilistic Engineering Mechanics},
volume = {74},
pages = {103485},
year = {2023},
author = {Konstantinos G. Papakonstantinou and Hamed Nikbakht and Elsayed Eshra},
}

@article{HELAL2026,
title = {Efficient rare event estimation for multimodal and high-dimensional system reliability via subset adaptive importance sampling},
journal = {Reliability Engineering \& System Safety},
volume = {272},
pages = {112473},
year = {2026},
author = {Sara Helal and Vìctor Elvira},
}

@article{GUO2025111078,
title = {Damage risk assessment of transmission towers based on the combined probability spatial-temporal distribution of strong winds and earthquakes},
journal = {Reliability Engineering \& System Safety},
volume = {261},
pages = {111078},
year = {2025},
author = {Xin Guo and Hongnan Li and Hao Zhang},
}

@article{luengo2020survey,
  title={A survey of Monte Carlo methods for parameter estimation},
  author={Luengo, David and Martino, Luca and Bugallo, M{\'o}nica and Elvira, V{\'\i}ctor and S{\"a}rkk{\"a}, Simo},
  journal={EURASIP Journal on Advances in Signal Processing},
  volume={2020},
  number={1},
  pages={25},
  year={2020},
  publisher={Springer}
}

@article{cornuet2012adaptive,
  title={Adaptive multiple importance sampling},
  author={CORNUET, JEAN-MARIE and MARIN, JEAN-MICHEL and Mira, Antonietta and Robert, Christian P},
  journal={Scandinavian Journal of Statistics},
  volume={39},
  number={4},
  pages={798--812},
  year={2012},
  publisher={Wiley Online Library}
}

@article{Moral2006,
    author = {Del Moral, Pierre and Doucet, Arnaud and Jasra, Ajay},
    title = {Sequential Monte Carlo Samplers},
    journal = {Journal of the Royal Statistical Society Series B: Statistical Methodology},
    volume = {68},
    number = {3},
    pages = {411-436},
    year = {2006},
    month = {06},
}

@article{wang2021hermite,
  title={Hermite polynomial normal transformation for structural reliability analysis},
  author={Wang, Jinsheng and Aldosary, Muhannad and Cen, Song and Li, Chenfeng},
  journal={Engineering Computations},
  volume={38},
  number={8},
  pages={3193--3218},
  year={2021},
  publisher={Emerald Publishing Limited}
}

@article{tong2021normal,
  title={Normal transformation for correlated random variables based on L-moments and its application in reliability engineering},
  author={Tong, Ming-Na and Zhao, Yan-Gang and Lu, Zhao-Hui},
  journal={Reliability Engineering \& System Safety},
  volume={207},
  pages={107334},
  year={2021},
  publisher={Elsevier}
}

@article{wang2025structural,
  title={Structural reliability assessment using quartic normal transformation},
  author={Wang, Tianfeng and Ji, Xiaowen and Zhao, Yan-Gang},
  journal={Scientific Reports},
  volume={15},
  number={1},
  pages={990},
  year={2025},
  publisher={Nature Publishing Group UK London}
}

\end{document}